\documentclass[aps,tightenlines,showpacs]{revtex4}
\usepackage{graphicx}
\usepackage{float}
\begin{document}

%

\newcommand\dd{\mbox{d}}   
\renewcommand\exp{\mbox{\rm exp}}  
\newcommand\tr{\mbox{\rm tr}} 
\newcommand\eff{{\rm eff}}       

\newcommand\intks {\it \frac{\dd^4 \K}{(2\pi)^4}}   
\newcommand\intps{\int \frac{\dd^4 \P}{(2\pi)^4}}
\newcommand\intxs{\int d^3 \X}
\newcommand\intys{\int d^3 \Y} 

\newcommand{\xx}{\noindent}
\newcommand{\vs}{\vspace*{.5cm}}
\newcommand\ra{\rightarrow}
\newcommand\la{\leftarrow}
\newcommand\lla{\longleftarrow}
\newcommand\lra{\longrightarrow}
\newcommand\llra{\longleftrightarrow}

\newcommand{\lan}{\langle}     
\newcommand{\ran}{\rangle}     

\newcommand\ot{\otimes}        

\newcommand\Del{\mbox{\boldmath $\partial$}}
\newcommand\del{\partial}

\newcommand\ncdot{\!\cdot\!}   

\newcommand{\fxf}{\!\!\!\!}      
\newcommand{\fxt}{\!\!\!}        
\newcommand{\fxd}{\!\!}          

\newcommand\be{\begin{equation}}
\newcommand\ee{\end{equation}}
\newcommand\bea{\begin{eqnarray}}
\newcommand\eea{\end{eqnarray}}

\newcommand\ba{\begin{array}}
\newcommand\ea{\end{array}}


\title {Scattering amplitudes at finite temperature}

\author{ M.E. Carrington${}^{a,b}$, Hou Defu${}^{c,d}$ and R. Kobes${}^{b,d}$}
\email{meg@theory.uwinnipeg.ca; hdf@theory.uwinnipeg.ca;
randy@theory.uwinnipeg.ca}
 \affiliation{ ${}^a$ Department of Physics, Brandon University, Brandon,
Manitoba,
R7A 6A9 Canada\\
 ${}^b$  Winnipeg Institute for Theoretical Physics, Winnipeg, Manitoba \\
${}^c$ Institute of Particle Physics, Huazhong Normal University, 430070
Wuhan,
China \\
${}^d$ University of Winnipeg, Winnipeg, Manitoba, R3B 2E9 Canada }

\begin{abstract} 
We present a simple set of rules for obtaining the imaginary part of a self
energy diagram at finite temperature in terms of diagrams that
correspond to physical scattering amplitudes. 
\end{abstract}
\maketitle

\section{Introduction}
\label{s:intro}

In this paper we discuss a set of rules for calculating the imaginary part of
self energy diagrams. These rules have a simple diagrammatic representation in terms
of scattering amplitudes. They have been
deduced by studying a large set of diagrams; a derivation 
from first principles is in
progress and will be presented in another paper.

It is well known that the imaginary part of the retarded self energy is an
important quantity in thermal field theory: it provides information about
decay and production rates of particles, among other things. The physics that
is contained in the imaginary part of the self energy is revealed by writing
it as the product of two scattering amplitudes.
At the one loop level, the structure of the scattering amplitude has been
understood for some time \cite{Weld}. Extension to higher loops is not
straightforward. In this paper we discuss cutting rules in the context of
this problem.  We show that rules exist that
make it easy to understand the physical
content of the imaginary part of a self energy diagram at high loop order. 

We begin by reviewing some basic concepts and defining some notation: 

\xx {\bf a)} Off shell propagators describe the behaviour of fields that
propagate through the medium until they undergo interactions with other fields.
Diagrammatically, off shell propagators are represented by continuous lines 
that begin and end at interaction vertices. 

\xx {\bf b)} On shell propagators carry delta functions of the form
$\delta(P^2-m^2)$ where $P$ is the momentum of the field. On shell propagators
correspond to fields that do not propagate through the medium. 

\xx {\bf c)} Consider any closed loop in which all of the propagators are off
shell and momentum is free to flow around the loop. If any one of the
propagators in the loop is put on shell, momentum is no longer free to flow
around the loop and the loop is effectively `opened.'

\xx {\bf d)} There are two kinds of propagators that are on shell. We call
these two types of propagators ``cut'' propagators and ``tic-ed'' propagators.
Cut propagators and tic-ed propagators carry different thermal factors. This
point will be explained in detail. 

\xx {\bf e)} A ``cut line'' is a line that divides the self energy into two
pieces, each of which has one external leg. Any propagator that is crossed by a
cut line is put on shell and becomes a ``cut'' propagator. 

\xx {\bf f)} Diagrammatically, our notation is as follows. In a self energy
diagram, a cut propagator is a propagator that is crossed by the cut line and a
tic-ed propagator is drawn with a double tic mark through it. To obtain
scattering amplitudes, all on shell propagators (cut or tic-ed) are split into
two pieces, each of which has an end that is not connected to a vertex. As a
result, the cut line divides the self energy diagram into two separate
amplitudes, and (as will be explained below) the tic-ed propagators cause each
amplitude to have the form of a tree amplitude, with no closed loops. 
The lines obtained from the splitting of on shell propagators represent the
emission or absorption of fields by the medium. The lines that represent
absorbed fields  are drawn slanting backwards from the vertex and lines that
represent emitted particles slanting forward from the vertex. 

\xx {\bf g)} For any given diagram, the number of $\delta$ functions, or the
number of on shell~=~(cut~+~tic-ed) propagators, is equal to $L+1$ where $L$ is
the number of loops.

To begin, we consider a one loop calculation. Following Weldon \cite{Weld} we
look at the simple case of a scalar field $\Phi$ coupled to two other scalars
$\phi_1$ and $\phi_2$ through a cubic interaction. The production rate for the
field $\Phi$ is obtained from the imaginary part of the one loop self energy
shown in Fig.[1a]. We obtain the imaginary part by drawing a cut line through
the diagram. At one loop, there is only one way to draw a cut line so that each
half of the diagram contains one of the two external legs (Fig.[1b]). This cut
line produces two cut propagators, each of which carries a delta function
(there are no tic-ed propagators in this case). 
\par\begin{figure}[H]
\begin{center}
\includegraphics[width=7cm]{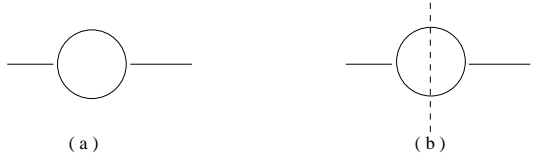}
\end{center}
\caption{  One-loop
self-energy}
 \label{fgm1}
\end{figure}
Expanding these two delta functions we find that the imaginary part of this
diagram can be written as the sum of four terms, each of which corresponds to
the square of a scattering amplitude \cite{Weld}.  We obtain,
\bea
{\rm Im} \Pi(q_0,\vec q) = -g^2\int
\frac{d^3p}{(2\pi)^3}&&\frac{2\pi}{2\omega_1
2\omega_2}[\delta(q_0-\omega_1-\omega_2)[(1+n_1)(1+n_2)-n_1n_2]\nonumber \\
&&~~~~+
\delta(q_0+\omega_1-\omega_2)[(n_1(1+n_2)-(1+n_1)n_2]\nonumber \\
&&~~~~+
\delta(q_0-\omega_1+\omega_2)[(1+n_1)n_2-n_1(1+n_2)]\nonumber \\
&&~~~~+
\delta(q_0+\omega_1+\omega_2)[n_1 n_2-(1+ n_1)(1+n_2)]]\label{weld}
\eea
where $\omega_1 = \sqrt{\vec p^2+m_1^2}$, $\omega_2 = \sqrt{(\vec p + \vec q)^2
+ m_2^2}$ and $n_1 = 1/(e^{\beta \omega_1}-1)$ with $\beta$ the inverse
temperature and $g$ the coupling constant.
The first term in this expression corresponds to the probability for the decay
$\Phi\rightarrow \phi_1\phi_2$ with a statistical weight $(1+n_1)(1+n_2)$ for
stimulated emission, minus the probability for the inverse decay
$\phi_1\phi_2\rightarrow\Phi$ with the weight $n_1 n_2$ for absorption. Note that the
thermal factor $(1+n_1)(1+n_2)-n_1n_2$ which reflects the physics of the
process involved could be written in the mathematically simpler form
$1+n_1+n_2$ at the cost of loosing information about the physics. 
Similarly, the second term gives the probability for the decay $\Phi
\phi_1\rightarrow \phi_2$ (which involves the absorption of a $\phi_1$ field
and the emission of a $\phi_2$ field), minus the probability for the reverse
process $\phi_2\rightarrow\Phi\phi_1$, with appropriate thermal weights. The
interpretation of the third and fourth terms is straightforward. These four
process are shown in Fig.[2].
\par\begin{figure}[H]
\begin{center}
\includegraphics[width=7cm]{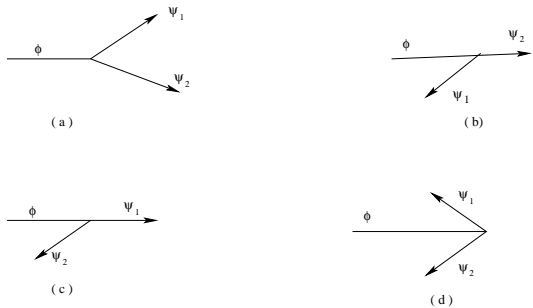}
\end{center}
\caption{  The scattering amplitudes corresponding to Fig.[\ref{fgm1}]}
\label{fgn2}
\end{figure}

We would like to study what happens at higher loops where scattering amplitudes
have a  much more complicated structure.  The task of
calculating the imaginary part of the self energy and separating it into
scattering amplitudes is complicated in different ways, depending on the
technique that is used to calculate the imaginary part of the self energy.

In the imaginary time formalism (ITF), one starts in Euclidean space,
calculates Green functions with imaginary time arguments, and performs an
analytic continuation to real time at the end of the calculation.
One attractive feature of the ITF is that it satisfies the intuitive belief
that it should be possible to obtain  finite temperature field theory from zero
temperature field theory by adding thermal weights to the Feynman rules,  in
some fashion.  The ITF is mathematically simpler than the RTF because of this
structure. However, the price one pays for mathematical simplicity is that some
physical processes are hidden.  Thinking more carefully we realize that
this feature of the ITF is not surprising since we should not in fact expect
that finite temperature field theory should have the same structure as zero
temperature field theory. In a thermal situation, individual fields do not
simply propagate through the vacuum, but interact with a medium. Consequently,
a specific scattering amplitude will involve a number of interference process
that are not present at zero temperature.  All of these processes are, of
course, present in the ITF calculation, but the compactness of the notation
effectively hides them from view. A lengthy procedure for extracting physical
amplitudes from the ITF has been discussed by Wong \cite{Wong}. This extremely
complicated calculation has been carried out explicitly by Majumder and Gale
for the two loop vector boson self energy in thermal QCD \cite{Gale}.

In the real time formalism (RTF), one works in Minkowski space and Green
functions with real time arguments are obtained directly. It is well known that
the RTF is mathematically more complicated than the ITF because of the doubling
of field degrees of freedom. In the closed time path (CTP) representation of the RTF, the contour has two
branches: the top one (${\cal C}_1$) runs from negative infinity to positive
infinity, and the bottom one (${\cal C}_2$) runs backwards in the other
direction. These two branches give the propagator a matrix structure. The four
components of the 2$\times$2 matrix are labeled $D_{11}$, $D_{12}$, $D_{21}$,
and $D_{22}$ and correspond to propagation along ${\cal C}_1$, propagation from
${\cal C}_1$ to ${\cal C}_2$, etc. The propagator $D_{11}$ corresponds to time
ordered propagation. The advantage of working in the RTF is that is it easier
to separate the imaginary part of a self energy into physical scattering
amplitudes. This point will be discussed in detail.

A great deal of work has already been done on the development of rules for the
calculation of the imaginary parts of diagrams in the RTF. One set of rules
was derived by Kobes and Semenoff \cite{KS} using 
the 1/2 representation of the RTF.
A set of rules based on more general representations,
such as the R/A or Keldysh representaions, has been developed by 
Gelis \cite{Gelis}. We have checked that these rules and
the ones discussed in this paper are equivalent, as they
must be. However, some intricate cancellations are required to extract the  scattering amplitudes from the rules of
\cite{KS,Gelis}.
Several other authors have looked directly at scattering
amplitudes. Aurenche and collaborators
have separated the scattering amplitudes contained the imaginary parts of the
two loop photon self energy diagrams by dividing the phase space of the
momentum integrals into regions that contain the different possible
combinations of signs of the frequencies of the fields \cite{Aurenche}.
However, this technique would be difficult to generalize at higher loops.
Brandt and collaborators have developed a diagrammatic representation for
retarded green functions in terms of tree scattering amplitudes in the high
temperature limit \cite{Brandt}. 

The paper is organized as follows. In section II we present the first part of
our cutting rules: how to determine the diagrams that contribute to the
imaginary part of a given self energy, and for each diagram, which propagators
are on shell, retarded, or advanced. Section III contains an discussion of
how to interpret the diagrams produced by our rules in terms of scattering
amplitudes. Section IV contains the second part of our rules: how to determine
the thermal factor for each propagator.
Section V contains a list of the self energy diagrams that were used to deduce
these rules, and the diagrams that contribute to their imaginary parts. In
section VI we present our conclusions.

\section{First part of the cutting rules: Propagators}

\subsection{Allowed Diagrams}

For any self energy diagram, draw all possible cut lines (a cut line is any
line that divides the diagram into two pieces, each of which contains an
external leg). A cut line that opens all loops will produce $L+1$ cut
propagators and thus $L+1$ delta functions.  No other propagators will be on
shell. Other cut lines will leave some loops unopened, and produce less than
the required number of $L+1$ delta functions. Add tic marks to these diagrams
in every possible way so that 1) all loops are open; 2)  the total number of
delta functions is equal to the $L+1$ and 3) it is possible to move from either
external leg, to the cut line by following a continuous path along uncut and
untic-ed propagators. As an example, for the scalar self energy in Fig.[3], the
allowed cut diagrams are shown in Fig.[4] (for a scalar theory, the diagrams in
Figs.[4a,b] and Figs. [4c,d] are equivalent). 
\par\begin{figure}[H]
\begin{center}
\includegraphics[width=7cm]{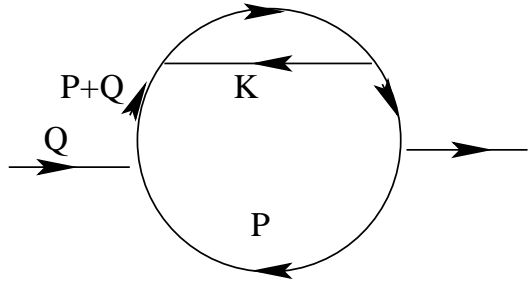}
\end{center}
\caption{A two-loop self energy  }
\label{fgm3}
\end{figure}
\par\begin{figure}[H]
\begin{center}
\includegraphics[width=7cm]{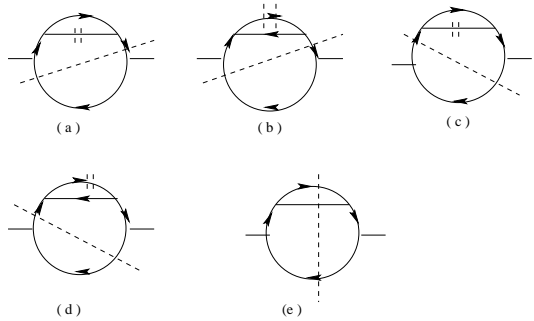}
\end{center}
\caption{The cuts that contribute to the imaginary part of Fig.[3]}
\label{fgm4}
\end{figure}
Two examples of diagrams that violate rule 3) are shown in Fig.[5]. Both of
these diagrams contain the right number of delta functions (5), and every loop
is opened, but it isn't possible to get from the right leg of either diagram to
the cut line without going through a cut or tic-ed propagator. Neither of these
diagrams should be drawn.
\par\begin{figure}[H]
\begin{center}
\includegraphics[width=8cm]{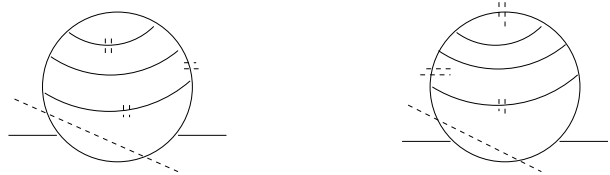}
\end{center}
\caption{Two diagrams that violate rule 3)}
\label{fgm5}
\end{figure}


\subsection {Uncut and Untic-ed Propagators }

Propagators that are uncut and untic-ed are either retarded or advanced.  There
is a rule to determine which.
Consider the imaginary part of a retarded self energy. Start from either
external leg and trace any continuous path through the diagram (continuous
means not traveling along a propagator that is cut or tic-ed). If the momentum
of a given propagator flows in the same direction as the path you are tracing,
the propagator is retarded. If the momentum flows in the direction opposite to
the direction of the path,  the propagator is advanced. This rule is
illustrated in Fig.[6].

\par\begin{figure}[H]
\begin{center}
\includegraphics[width=8cm]{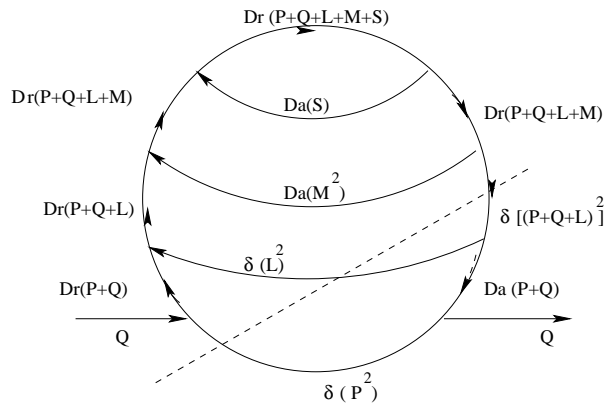}
\end{center}
\caption{An illustration of the propagator rule.}
\label{fgm6}
\end{figure}

\section{Scattering Amplitudes}

In this section, we discuss how to interpret the diagrams produced by our rules
as scattering amplitudes. 

\subsection{QED at One Loop}

Throughout most of this paper, we will work with $\phi^3$ theory.  This toy
theory allows us to avoid a lot of the mathematical complexities associated with
physical theories which are not directly relevant to the problem of how to
calculate the imaginary part of a self energy, and how to interpret the result
in terms of scattering amplitudes. However, in this section and the next, in
order to discuss the physics of scattering amplitudes in more familiar terms,
we will switch to QED. The structure of the scattering amplitudes in QED and
$\phi^3$ theory is similar since both theories have an interaction involving
three fields: $\phi^3$ theory has a cubic interaction, and QED has a
photon-electron-positron interaction. In the diagrams in this section and the
next, all arrows give the direction of flow of lepton number. We will look at
the photon self energy.

We start with the one loop diagram (Fig.[7a]). There is only one possible cut
line that separates the diagram into two pieces, so that each contains an
external leg. The amplitudes obtained from the right hand side are shown in
Fig.[7b] and have the same structure as those shown in Fig.[2]. As discussed in
the introduction, a fermion line that represents an emitted (absorbed) fermion
is connected to a vertex at only one end and is drawn slanting forwards
(backwards) from that vertex. Fig.[7b] shows the amplitude for a photon to
decay into an electron-positron pair $(\gamma\rightarrow e^+ e^-)$, the
amplitude for a photon to absorb a positron from the medium and emit an
electron $(\gamma e^+\rightarrow e^-$), etc. In the future, we restrict to
positive frequencies for the external field. In this case the process
represented by the last amplitude is kinematically forbidden. The left hand
side of the cut self energy in Fig.[7a] gives the conjugate of the amplitudes
in Fig.[7b], and thus the product of the right hand side and the left hand side
is a real probability.

\par\begin{figure}[H]
\begin{center}
\includegraphics[width=8cm]{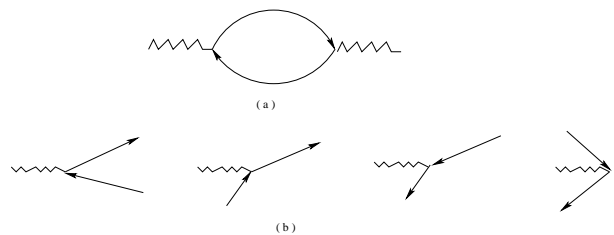}
\end{center}
\caption{One loop photon self energy and
scattering amplitudes.}
\label{fgm16}
\end{figure}
\subsection{QED at Two Loops}

There are two diagrams that contribute to the photon self energy at two loops.
They are shown in Figs.[8a,b].  For each of these diagrams, there is more than
one possible cut line.

\par\begin{figure}[H]
\begin{center}
\includegraphics[width=8cm]{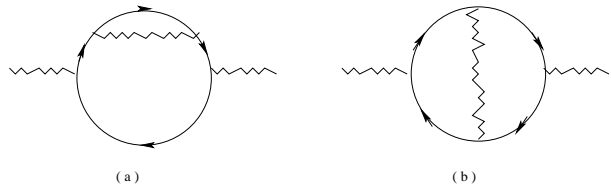}
\end{center}
\caption{Two loop photon self energies.}
\label{fgm17}
\end{figure}
\subsubsection{Central Cuts}

To begin with, for both diagrams, we look at the cut line that goes directly
through the center as shown in Fig.[9].
\par\begin{figure}[H]
\begin{center}
\includegraphics[width=8cm]{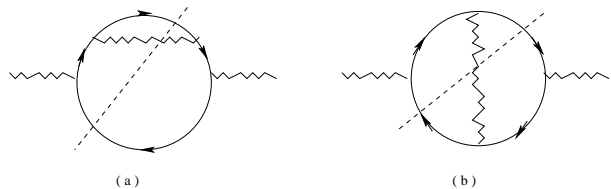}
\end{center}
\caption{Central cuts.}
\label{fgm18}
\end{figure}
In both cases, all loops are opened  and thus there are no tic-ed propagators.
Both cut lines cross one photon propagator and two fermion propagators. The
forward scattering amplitude is given by the terms in which all three particles
are emitted. The Compton scattering amplitude occurs when the photon is
emitted, and one fermion is emitted and one is absorbed. The amplitude for pair
production is produced when the photon is absorbed and both fermions are
emitted. When we restrict to positive external frequencies, no other
possibilities are kinematically allowed. We describe these processes below.

{\bf 1)} Forward Scattering

 The amplitude produced by left hand side of Fig.[9a] is shown in Fig.[10a].
The right hand side of the diagram is the conjugate 
amplitude. There is another contribution to the self energy that is the same as
Fig.[8a] except with the propagator correction on the lower line, or
alternatively, with the flow of lepton number in the fermion loop routed in the
opposite direction. This graph is just the hermitian transpose of Fig.[8a] and
the central cut produces the same amplitude with the roles of the electron and
positron reversed, as shown in Fig.[10b]. Again, the right hand side of the
diagram is the conjugate amplitude. These two diagrams
represent the amplitude for a photon to decay into an electron-positron pair
where one of the fermions emits an additional photon.

\par\begin{figure}[H]
\begin{center}
\includegraphics[width=8cm]{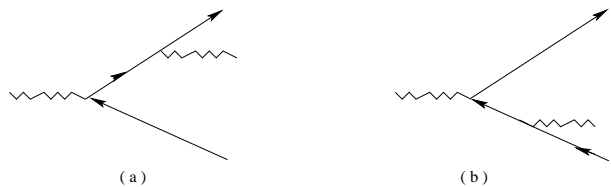}
\end{center}
\caption{Photon decay amplitudes from Fig.~[\ref{fgm18}a].}
\label{fgm19}
\end{figure}


The amplitude corresponding to the left hand side of Fig.[9b] is also shown in
Fig.[10a]. In this case however, the right hand side of the diagram does not
give the conjugate amplitude. Instead, we get the cross product of Fig. [10a]
with the conjugate of Fig.[10b].  By taking the cut line diagonally in the opposite direction,
we obtain the reversed cross product, with the electron and positron switched.
Combining these results we obtain the square of the amplitude shown in
Fig.[11], which is proportional to the photon decay probability.
\par\begin{figure}[H]
\begin{center}
\includegraphics[width=8cm]{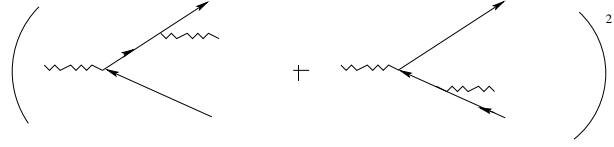}
\end{center}
\caption{Photon decay.}
\label{fgm20}
\end{figure}


{\bf 2)} Compton Scattering

We look at exactly the same cuts as above, but consider the amplitudes that
correspond to one emitted photon, one emitted fermion and
 one absorbed fermion.  The graph in Fig.[9a] produces the squares of the two
amplitudes shown in Fig.[12a].
 The diagram with the propagator correction on the bottom line gives the same
amplitudes, with the direction of lepton flow reversed.  
  The diagram in Fig.[9b] gives the two cross terms shown in Fig.[12b] .
\par\begin{figure}[H]
\begin{center}
\includegraphics[width=8cm]{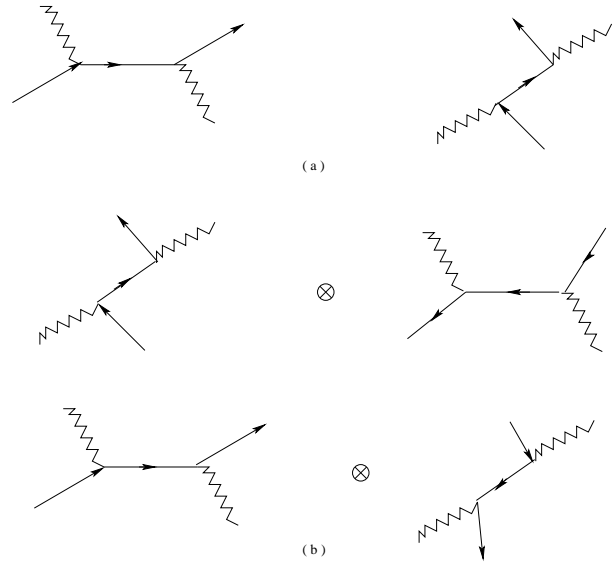}
\end{center}
\caption{Compton scattering amplitudes from Fig.~[\ref{fgm20}].}
\label{fgm21}
\end{figure}

The graphs that are the same as those in Fig.[12b] but with the fermion lines
reversed are produced by the cut on the opposite diagonal. Combining these
results we obtain the square of the amplitude shown in Fig.[13], which is
proportional to the Compton scattering rate. 
\par\begin{figure}[H]
\begin{center}
\includegraphics[width=8cm]{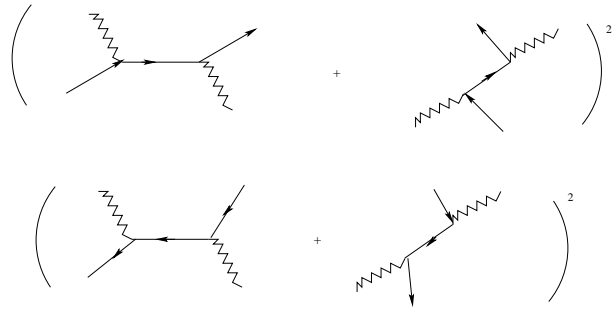}
\end{center}
\caption{Compton scattering.}
\label{fgm22}
\end{figure}


{\bf 3)} Pair Production

Following the same method, we obtain the square of the amplitudes for pair
production as shown in Fig.[14]. As before, the cross term is produced by
Fig.[9b].
\par\begin{figure}[H]
\begin{center}
\includegraphics[width=8cm]{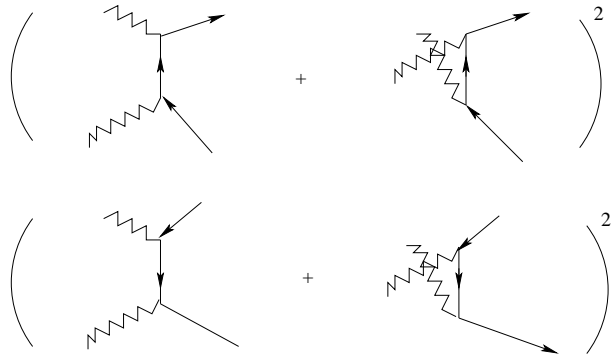}
\end{center}
\caption{Pair production.}
\label{fgm23}
\end{figure}

\subsubsection{Non-Central Cuts}

{}From this point on we will discuss only forward scattering amplitudes. Also,
we will not discuss the fact that each amplitude has an partner in which the
flow of fermion number is reversed. These graphs are obtained in exactly the
same way as was described above for the central cuts.

For both of the diagrams in Fig.[8] there is a cut that does not open both
loops. For the diagram in Fig.[8a] this cut leaves the loop formed by the
propagator correction unopened. There are two ways to open this loop with a
tic. These two graphs are shown in Fig.[15].
\par\begin{figure}[H]
\begin{center}
\includegraphics[width=8cm]{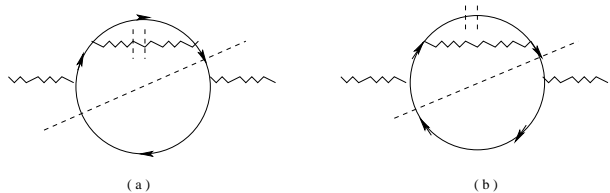}
\end{center}
\caption{Non-central cuts for Fig.[8a].}
\label{fgm24}
\end{figure}
The amplitudes corresponding to the left hand side of Fig.[15a] are shown in
Fig.[16a,b]. The first of these  graphs represents the amplitude for a photon
to decay into an electron-positron pair, where the electron subsequently
absorbs a photon from the medium and emits it back into the medium. The second
graph is the same as the first, except the photon is emitted into the medium
and then absorbed. This photon is sometimes referred to as a spectator field.
Spectators are on shell particles from the heat bath that enter with the
in-state and leave with the out-state without having interacted with the rest
of the participants. The right hand side of Fig.[15a] is shown in Fig.[16c].
The lone wiggly line represents the spectator particle from the heat bath. This
spectator field is normally not drawn, but its presence is necessary to obtain
an interference effect: when calculating the convolution of the two amplitudes,
it is necessary to have the same number of incoming and outgoing particles. We
can write a schematic equation to describe these amplitudes:
\bea
(\gamma \tilde \gamma \rightarrow e^+e^- \tilde \gamma) \otimes (\gamma
e^+e^-~;~~\tilde \gamma\rightarrow \tilde\gamma ) \nonumber
\eea
where the tilde indicates the spectator field.
\par\begin{figure}[H]
\begin{center}
\includegraphics[width=8cm]{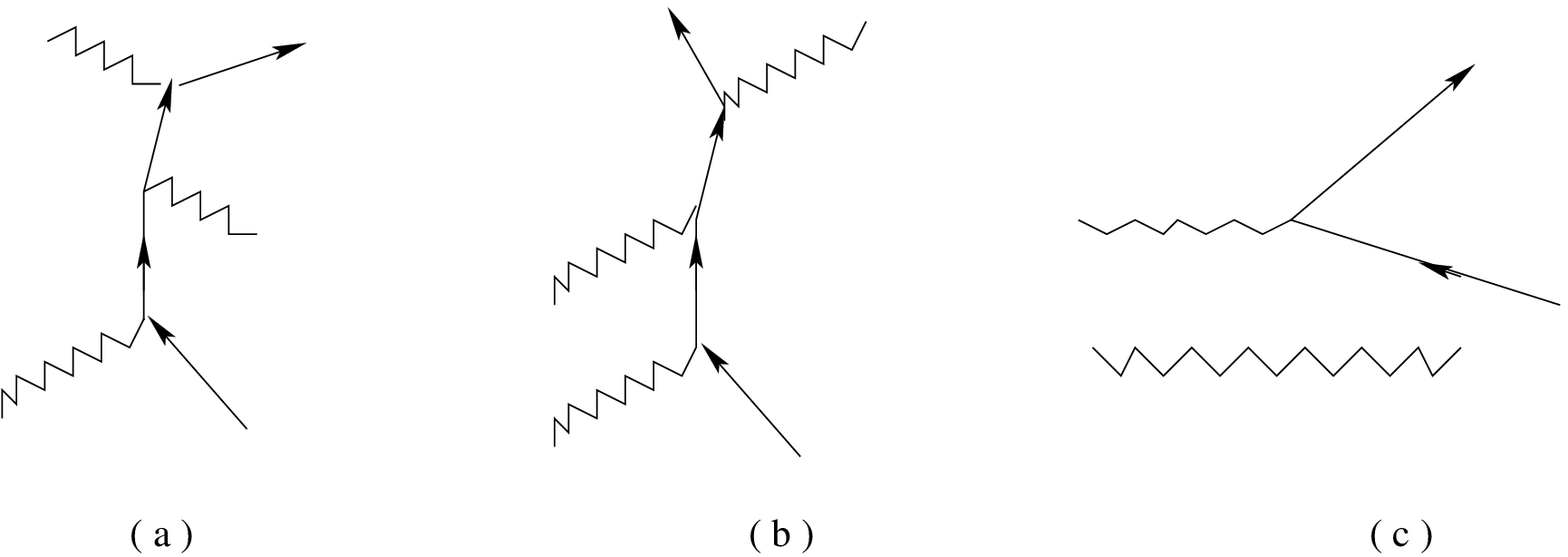}
\end{center}
\caption{Amplitudes from Fig.~[\ref{fgm24}a].}
\label{fgm25}
\end{figure}


The amplitudes from the left hand side of Fig.[15b] are shown in Fig.[17a-c].
Fig.[17a] represents the amplitude for a photon to decay to an
electron-positron pair, where the electron subsequently absorbs a positron from
the medium and emits it back into the medium. In Fig.[17b] this spectator
positron is emitted and then absorbed. Fig.[17c] shows the amplitude from the
right hand side of the self energy, with the spectator positron. Schematically
we write,
\bea
(\gamma \tilde e^+ \rightarrow e^+e^-\tilde e^+)\otimes (\gamma\rightarrow e^+
e^-~;~~\tilde e^+\rightarrow \tilde e^+) \nonumber
\eea

\par\begin{figure}[H]
\begin{center}
\includegraphics[width=8cm]{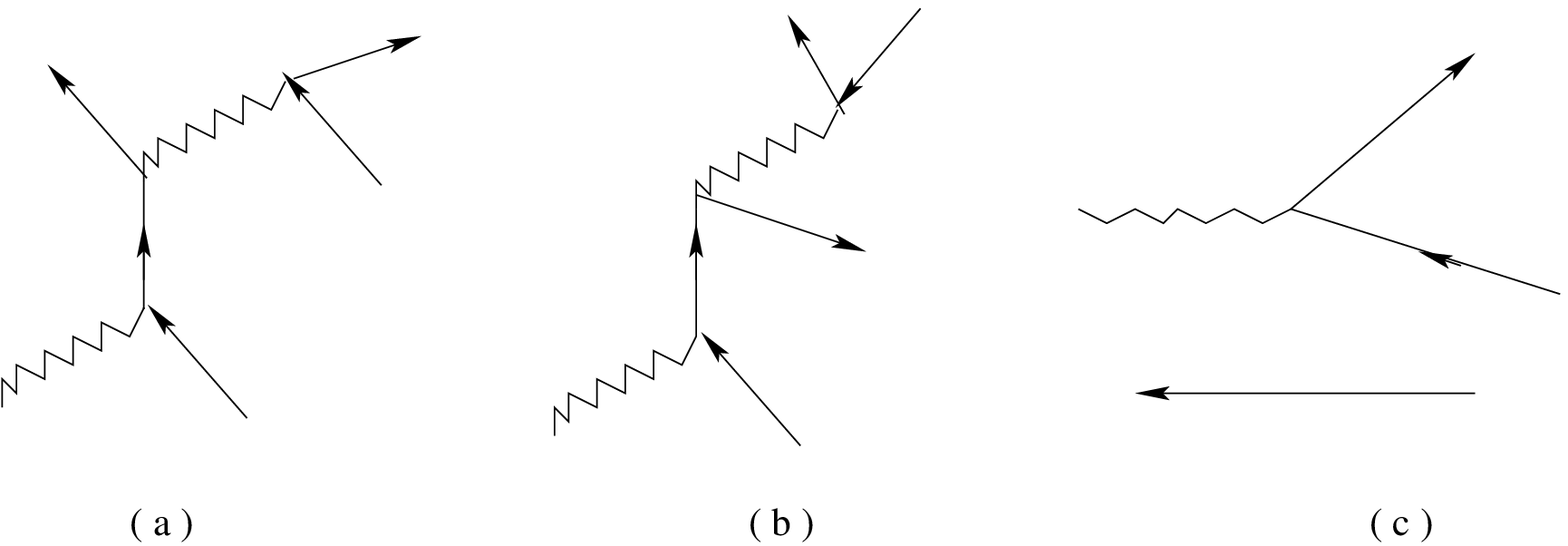}
\end{center}
\caption{Amplitudes from Fig.~[\ref{fgm24}b].}
\label{fgm26}
\end{figure}


A real probability is obtained by combining with the amplitudes produced by the
graphs that are just like those in Fig.[15], but with the cut line on the left
hand side. All of these graphs have been extracted from the imaginary time
calculation by the authors of \cite{Wong} and \cite{Gale}. In \cite{Wong} they
appear as Figs.[16,17] and in \cite{Gale} they are Fig.[12]. In the imaginary
time calculation, a great deal of tedious effort is required to separate these
physical contributions. Using our rules however, they appear immediately. 

Now we consider the non-central cut for Fig.[8b]. The unopened loop can be
opened with a tic on the photon line Fig.[18a], or a tic on the fermion line
Fig.[18b] (tic-ing the upper fermion line would change the spectator fermion to
a spectator anti-fermion). The amplitudes that result from Fig.[18a] are shown
in Fig.[19a-c]. The first two graphs show the decay of a photon into a
electron-positron pair with a spectator photon. In Fig.[19a] the electron
absorbs a photon from the medium and the positron emits a photon, and in
Fig.[19b] the electron emits a photon and the positron absorbs a photon.
Fig.[19c] shows the amplitude from the right hand side of the self energy in
Fig.[18a], with the spectator photon. We write schematically,
\bea
(\gamma \tilde \gamma \rightarrow e^+e^- \tilde\gamma)\otimes
(\gamma\rightarrow e^+e^-~;~~\tilde \gamma \rightarrow \tilde \gamma)
\nonumber
\eea
The amplitudes from Fig.[18b] are shown in Figs.[20a-c]. Schematically we
write,
\bea
(\gamma \tilde e^- \rightarrow e^+e^- \tilde e^-)\otimes (\gamma\rightarrow
e^+e^-~;~~ \tilde e^-\rightarrow \tilde e^-) \nonumber
\eea
\par\begin{figure}[H]
\begin{center}
\includegraphics[width=8cm]{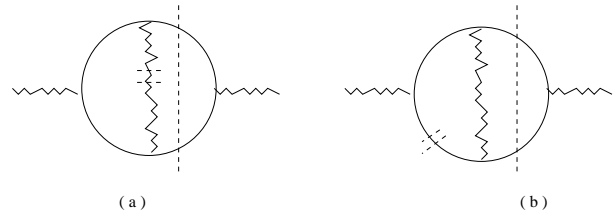}
\end{center}
\caption{Non-central cuts for Fig.[8b].}
\label{fgm27}
\end{figure}
\par\begin{figure}[H]
\begin{center}
\includegraphics[width=8cm]{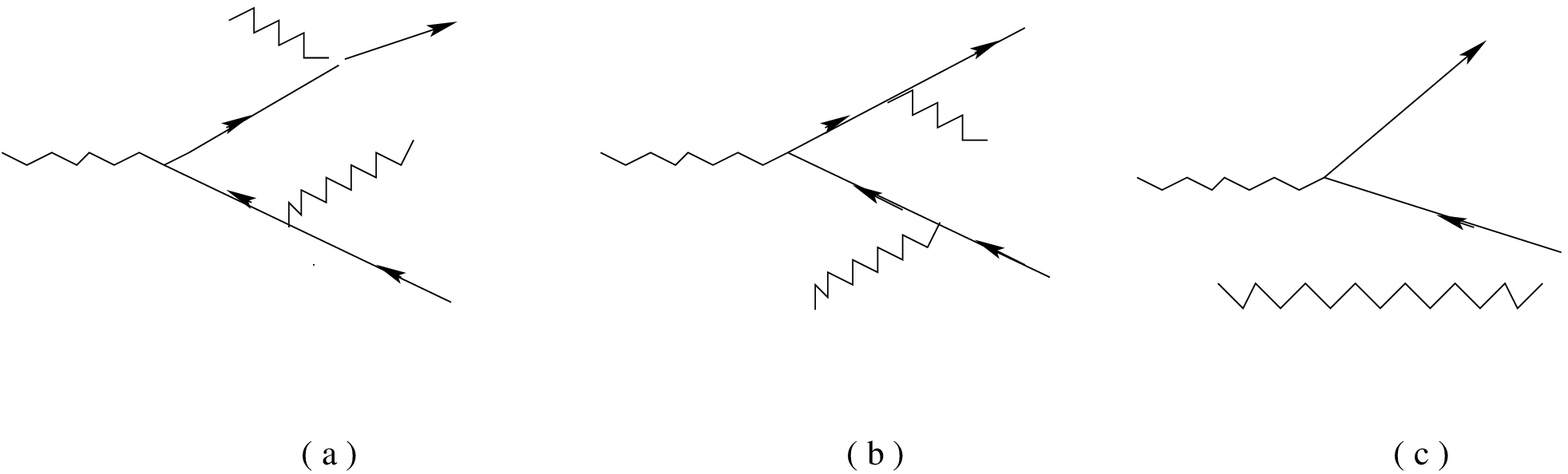}
\end{center}
\caption{Amplitudes from Fig.~[\ref{fgm27}a].}
\label{fgm28}
\end{figure}
\par\begin{figure}[H]
\begin{center}
\includegraphics[width=8cm]{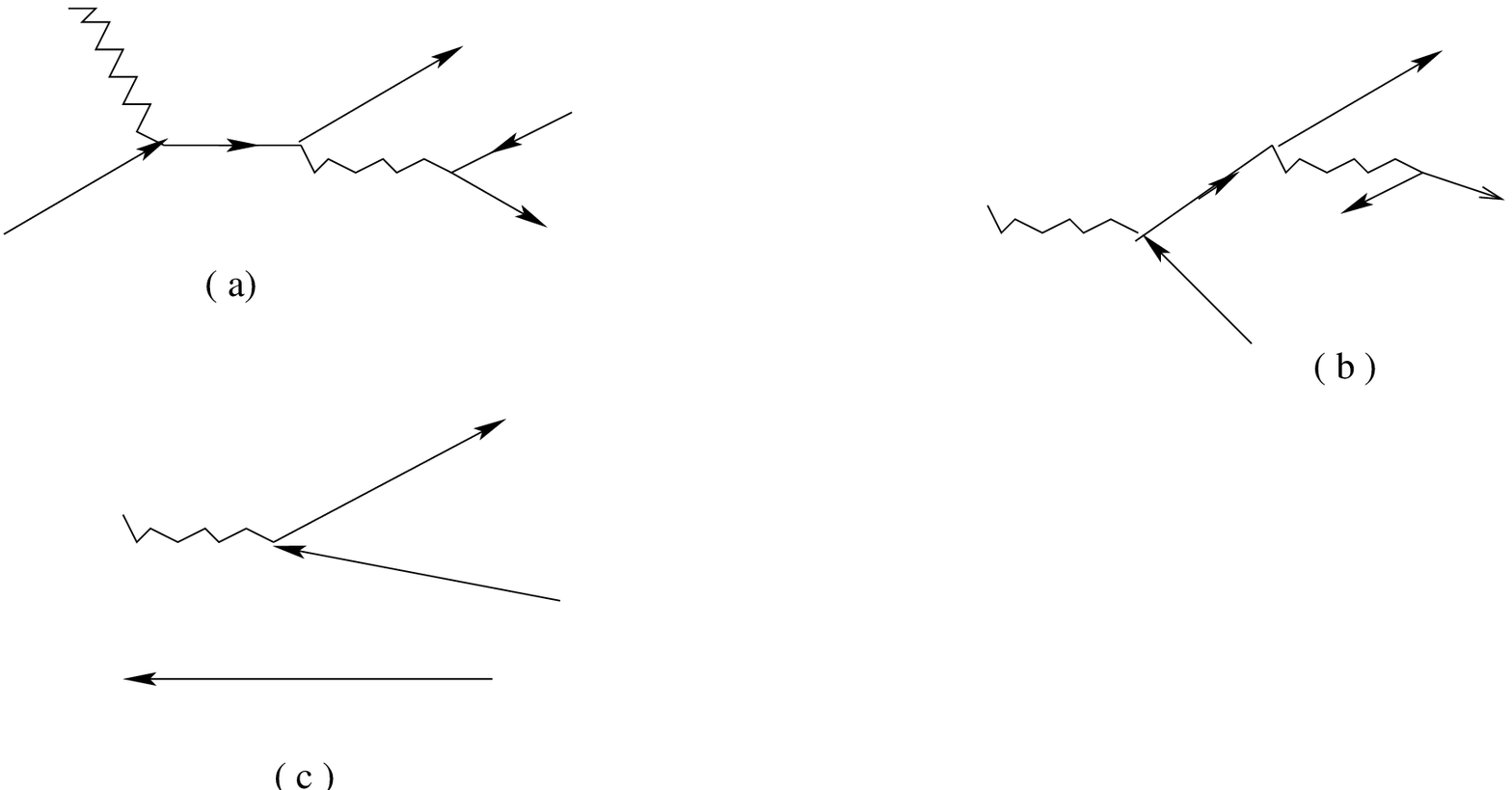}
\end{center}
\caption{Amplitudes from Fig.~[\ref{fgm27}b].}
\label{fgm29}
\end{figure}

These diagrams are Figs.[19,20] in \cite{Wong} and Figs.[10,11] in
\cite{Gale}.

\subsection{ $\phi^3$ Theory at More Than Two Loops}

In order to discuss scattering amplitudes at higher loop levels, we will
simplify the notation as follows. 

1) We revert to $\phi^3$ theory.

2) We will only draw the forward scattering amplitude (all cut fields on the
positive mass shell) associated with the left hand side of a given cut self
energy.

3) We simplify the diagrammatic representation of tic-ed fields as follows. All
tic-ed propagators correspond to fields that are emitted and then absorbed, or
absorbed and then emitted. Thus, if there are two tic-ed propagators on the
left hand side of a cut self energy, there are four different scattering
amplitudes corresponding to the four different ways in which these two fields
could be emitted and absorbed from the heat bath. Using the notation from the
first part of this section we would draw four diagrams, one for each of these
processes, with the tic-ed propagators represented by lines slanting forward or
backwards to represent emission or absorption respectively. In this subsection,
we will draw only one diagram instead of four, and represent emitted or
absorbed fields by arrows pointing straight up, or slightly slanted to one side
or the other in the case where two fields are emitted/absorbed from the same
vertex and could not be seen on the figure if they were drawn on top of each
other.

4) All propagators will be drawn as lines slanting up to the right. Note that if
we define momentum variables so that momentum flows to the right, all
propagators are retarded (following the rule in section IIB), and thus the
scattering amplitude has a causal interpretation: time flows to the right along
any continuous line.

5) Cut propagators represent fields that are emitted and absorbed on opposite
sides of the diagram. These cut lines are drawn as horizontal arrows pointing
to the right. 

An example of this notation is shown in Fig.[21]. In the next section we
describe in detail how to determine the appropriate thermal factor for a given
cut diagram by looking at the corresponding scattering amplitude.
\par\begin{figure}[H]
\begin{center}
\includegraphics[width=8cm]{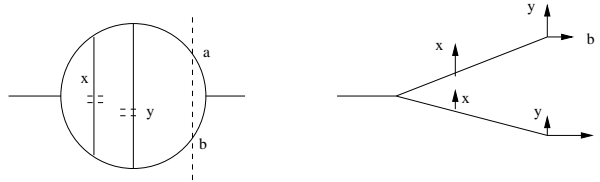}
\end{center}
\caption{An example of the notation to be used in section IV.}
\label{fgmn21}
\end{figure}

\section{Second Part of the Cutting Rules: Thermal Factors}

In the introduction we said that both cut propagators and tic-ed propagators
are on shell, and that the difference between them is that they carry different
thermal factors. The key to determining these thermal factors is understanding
the physical role of the cut and tic-ed propagators. We will discuss cut and
tic-ed propagators separately.

\subsection{Cut Propagators}

Return for a moment to the one loop example discussed in the introduction.
As discussed previously, the first term in Eqn.(\ref{weld}) comes from the
square of the forward scattering amplitude and  represents the probability for
the decay $\Phi\rightarrow \phi_1\phi_2$, minus the probability for the inverse
decay. The field $\Phi$ is the field whose self energy we are calculating. The fields $\phi_1$ and
$\phi_2$ are the fields produced by the decay of the $\Phi$ field (or the
fields which combine to produce a $\Phi$ field) and they appear as emitted
lines on the right side of the scattering amplitude.  This interpretation of
the physical role played by the fields corresponding to the cut propagators
allows us to understand the associated thermal factors.
The thermal factor $(1+n_1)(1+n_2)-n_1 n_2$ is the statistical weight
associated with the probability for the process $\Phi\rightarrow \phi_1\phi_2$
minus the statistical weight for the inverse process.

The generalization of this idea is straightforward. 
We write the thermal factor for the cut propagators as an equation of
the form,
\bea
\Pi_i \Pi_j [(1+n_i)n_j-n_i(1+n_j)] \label{cutlines}
\eea
where the product over $i$ multiplies over fields whose momenta approach the cut  line
from the left, and the product over $j$ multiplies over fields whose momenta  approach the
cut  line from the right. 
It is easy to understand the structure of this expression. 
Performing the loop integrals puts each cut propagator on either the positive or negative mass shell. A particle that approaches the cut from the left and is on the positive mass shell carries a factor $(1+n(\omega))$ in the first term of Eqn.(\ref{cutlines}), and corresponds to an emitted particle. If the same particle were on the negative mass shell, the corresponding thermal factor would be $(1+n(-\omega)) = -n(\omega)$, which correctly reflects the fact that the emission of a positive energy particle is equivalent to the absorption of a negative energy paarticle.  
Note also that the momentum of a cut
propagator could flow in either direction (right or left), depending on the
definitions chosen for the momentum variables in the loops. Of course, physics
must not depend on the choice of an integration variable. The identity $1+n(-x) = -n(x)$ ensures that all definitions of loop monenta are equivalent.

\subsection{Tic-ed Propagators}

Now we consider tic-ed propagators. Both cut and tic-ed propagators are on
shell, and both correspond to fields that are emitted and absorbed. However,
the fields associated with tic-ed propagators are  emitted and absorbed on the
same side of the cut self energy diagram. They represent interactions with
fields in the heat bath.  Naively, it appears that a tic-ed propagator
should carry a factor $N: = (1+n) + n$ which would give the appropriate
statistical weight for a field that is emitted and then subsequently absorbed.
(Of course the same factor appears if emission and absorption occur in the
opposite order). For convenience, we include a factor of $1/2$ (which will be
explained below) and use a thermal factor of the form
\bea
 N/2 \label{ticruleA}
 \eea 
In most cases this naive guess gives the correct answer.  Note that the
interaction represented by the thermal factor $N/2$ is essentially trivial: the
particle is produced and disappears without undergoing any interactions with
the rest of the system. As discussed in section III, these particles are
sometimes called `spectators.'

There are some cases in which the fields associated with the tic-ed propagators
undergo non-trivial interactions with the system, and in these cases the naive
thermal factor is incorrect. These diagrams always have more than one tic-ed
propagator, and thus only occur for self energies with three or more loops. The
authors of \cite{Wong,Gale} have not gone beyond the two loop level and thus
have not seen this effect. In order to identify the cut self energy diagrams in
which these non-trivial interactions occur, and to determine the appropriate
thermal factor, we need to look at the corresponding scattering amplitudes.
We describe below the rule for determining if the fields represented by the
tic-ed fields undergo non-trivial interactions, and for obtaining the thermal
factors for these fields.  We illustrate these rules by looking at several cuts
of the diagrams shown in Fig.[22].

\par\begin{figure}[H]
\begin{center}
\includegraphics[width=8cm]{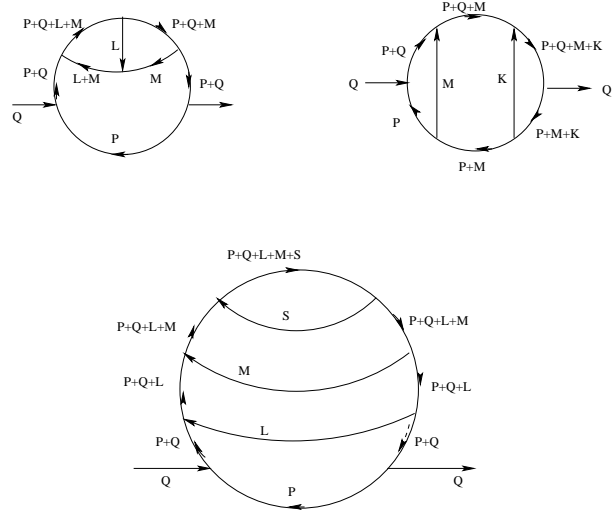}
\end{center}
\caption{Diagrams used to illustrate the rule in section IV-B2.}
\label{fgmm22}
\end{figure}

\subsubsection{Notation}

We define two kinds of tic-ed propagators.

For type-A tic-ed propagators, in the corresponding scattering amplitude, the
emitted-absorbed pair of fields from the tic-ed propagator are `unsplit' by either of the emitted/absorbed fields associated with the cut line.

For type-B tic-ed propagators, in the corresponding scattering amplitude, the emitted/absorbed pair of fields from the tic-ed propagator are `split' by one of the emitted/absorbed fields associated with the cut line.

The meaning of the expressions `split' and `unsplit' is explained below. The cut propagators are labeled $a$ and  $b$ in this section. The emitted/absorbed lines in the scattering amplitude that are associated with the cut propagators are horizontal arrows labeled  $a$ and  $b$.  The tic-ed propagators are labeled $\{x,~y,~z~\cdots\}$. The corresponding emitted/absorbed fields in the scattering amplitude appear as pairs labeled $\{x,~y,~z~\cdots\}$ (see Fig.[21]). Consider a tic-ed propagator labeled $x$. In order to determine if this is a type-A or type-B tic-ed propagator, one proceeds as follows. Trace a path through the scattering amplitude from a to b. If you encounter both or neither of the fields labeled $x$, than the tic-ed propagator labeled $x$ in the corresponding self energy is `unsplit' (type-A). If you encounter one $x$ field, than the tic-ed propapator is `split' (type-B). Numerous examples of this notation are given below.

\subsubsection{The Rule}

{\bf a)} If the scattering amplitude corresponds to a cut self energy that
contains an arbitrary number of type-A tic-ed propagators, all of these
propagators carry the naive thermal factor $N$. Several examples are shown in
Fig.[23]. The thermal factors for these four diagrams are respectively, \bea
&&\frac{1}{4}((1+n_{p+q})n_p-(1+n_p)n_{p+q})N(m)N(l+m)\nonumber \\
&&\frac{1}{4}((1+n_{p+q})n_p-(1+n_p)n_{p+q})N(l)N(p+q+m)\nonumber\\
&&\frac{1}{4}((1+n_{p+q+m+k})n_{p+m+k}-(1+n_{p+m+k})n_{p+q+m+k})N(p)N(k)
\nonumber \\
&&\frac{1}{4}((1+n_{p+q})n_p-(1+n_p)n_{p+q})N(p+q+m+l+s)N(p+q+l+m)N(l)
\nonumber\eea



\par\begin{figure}[H]
\begin{center}
\includegraphics[width=8cm]{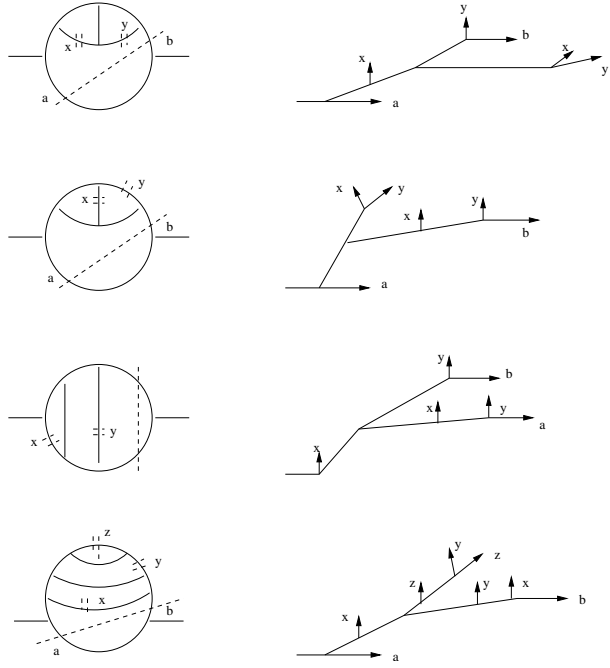}
\end{center}
\caption{
Self energies with type-A tic-ed propagators and the corresponding scattering
amplitudes }
\label{fgmn22}
\end{figure}

{\bf b)} If there is one type-B tic-ed propagator and an arbitrary number of
type-A tic-ed propagators, all propagators carry the naive thermal factor.
Several examples are shown in Fig.[24]. In the first two diagrams, the tic-ed
propagator marked $y$ is type-B, in the third diagram the tic-ed propagator
marked $x$ is type-B. The thermal factors for the three diagrams are
respectively,
\bea
&&\frac{1}{4}((1+n_{p+q+m+k})n_{p+m+k}-n_{p+q+m+k}(1+n_{p+m+k}))N(p)N(p+m)
\nonumber \\
&&\frac{1}{4}((1+n_{p+q+m+k})n_{p+m+k}-n_{p+q+m+k}(1+n_{p+m+k}))N(p)N(p+q+m)
\nonumber \\
&&\frac{1}{8}((1+n_{p+q})n_{p}-n_{p+q}(1+n_{p}))N(m)N(p+q+l)N(p+q+m+l+s)
\nonumber
\eea

\par\begin{figure}[H]
\begin{center}
\includegraphics[width=8cm]{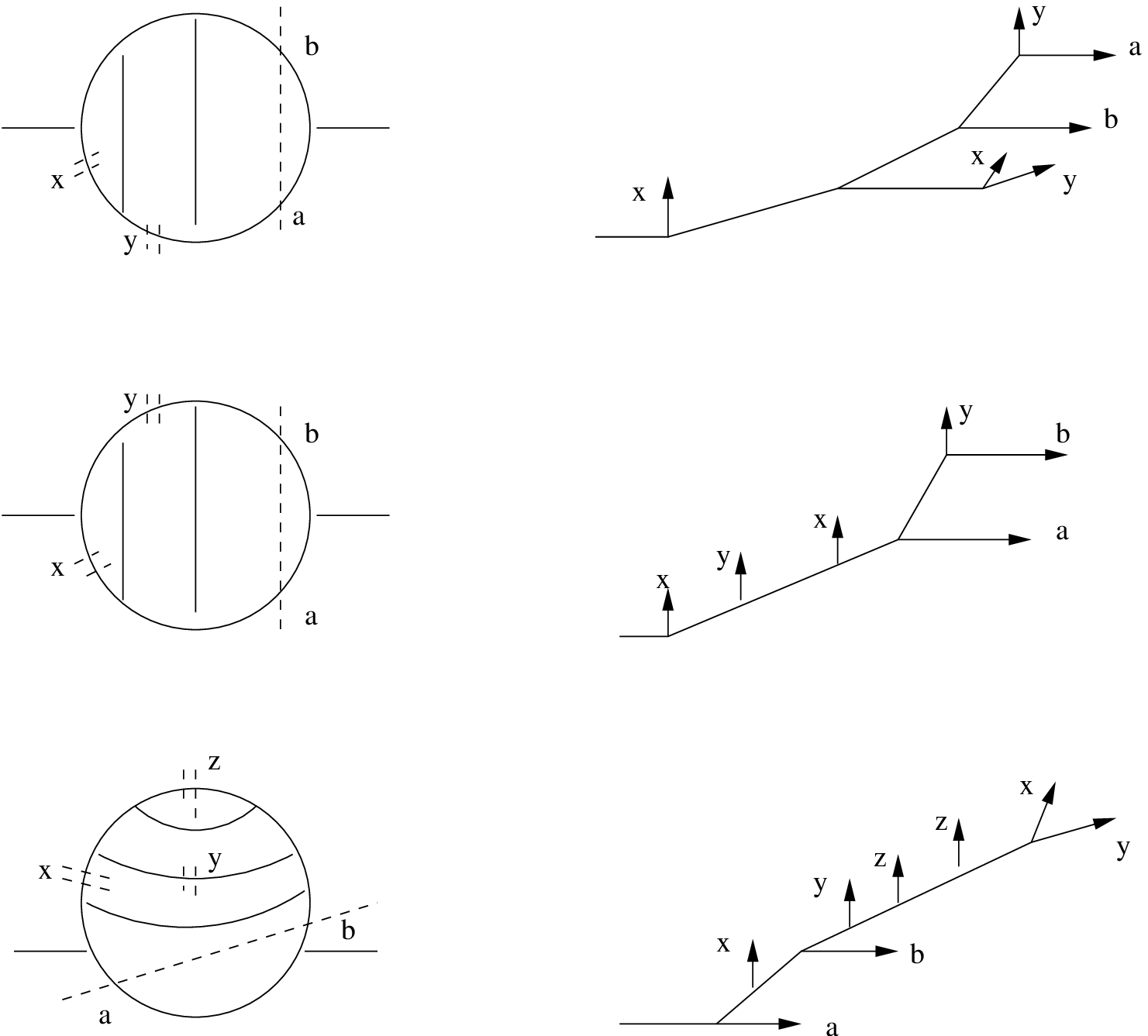}
\end{center}
\caption{Self energies with  one type-B  ticed propagator and  the
 corresponding scattering  amplitudes.}
\label{fgmn23}
\end{figure}

{\bf c)} If the scattering amplitude corresponds to a cut self energy with more
than one type-B tic-ed propagator, the thermal factors for those propagators
will not have the naive form. The correct thermal factor can be written,
\bea
2\Pi_i \Pi_j [(1+n_i)n_j+n_i(1+n_j)] \label{ticrule}
\eea
The notation is described below. Look at the set of fields coming from tic-ed propagators that appear between $a$ and $b$. Identify the vertex in the self energy that these fields came from. The product $i$ runs over fields whose momentum flows
away from this vertex, and the product $j$ runs over fields whose momentum flows
towards this vertex.
(Note that since the factor above is symmetric under interchange of $i$ and $j$
this definition could be reversed).
This factor looks almost exactly like the factor for cut propagators, expect
for the relative plus sign. This point and the factor of 2 in (\ref{ticrule})
will be explained below. Several examples are shown in Fig.[25]. For the first
diagram, the tic-ed propagators marked $x$ and $y$ are both type-B. The $x$ in the centre of the scattering amplitude came from a vertex that the momentum
of the tic-ed propagator flowed towards, and the $y$ in the centre of the
scattering amplitude came from a vertex that the momentum of the tic-ed
propagator flowed away from. Thus, the thermal factor for these two propagators
is $2((1+n_m)n_{l+m+p+q} + n_m (1+n_{l+m+p+q}))$, and the full thermal factor
is
\bea
((1+n_{p+q})n_p-n_{p+q}(1+n_p))N(p+q+m+l+s) ((1+n_m)n_{l+m+p+q} + n_m
(1+n_{l+m+p+q})) \nonumber 
\eea
In the second diagram, all three tic-ed propagators are type-B. In the centre of the scattering amplitude, the propagators marked $y$ and $z$ have
momenta that flow towards the vertex, and the propagator marked $x$ has
momentum that flows away from the vertex. The full thermal factor is
\bea
2((1+n_{p+q})n_p-n_{p+q}(1+n_p))((1+n_{p+q+m+l+s})n_m n_s +
n_{p+q+m+l+s}(1+n_m)(1+n_s)) \nonumber
\eea
The thermal factors for the last four diagrams in Fig.[25] are, in order,
\bea
&&2((1+n_{p+q+m+k})n_{p+m+k}-n_{p+q+m+k}(1+n_{p+m+k}))((1+n_m)(1+n_p)+n_m n_p)
\nonumber \\
&&2((1+n_{p+q+m+k})n_{p+m+k}-n_{p+q+m+k}(1+n_{p+m+k}))((1+n_m)(1+n_{p+q})+n_m
n_{p+q}) \nonumber \\
&&2((1+n_{p+q})n_p-n_{p+q}(1+n_p))((1+n_l)n_{p+q+m+l}+n_l(1+n_{p+q+m+l}))
\nonumber \\
&&2((1+n_{p+q})n_p-n_{p+q}(1+n_p))((1+n_l)n_{m+l}+n_l(1+n_{m+l}))\nonumber
\eea

\par\begin{figure}[H]
\begin{center}
\includegraphics[width=8cm]{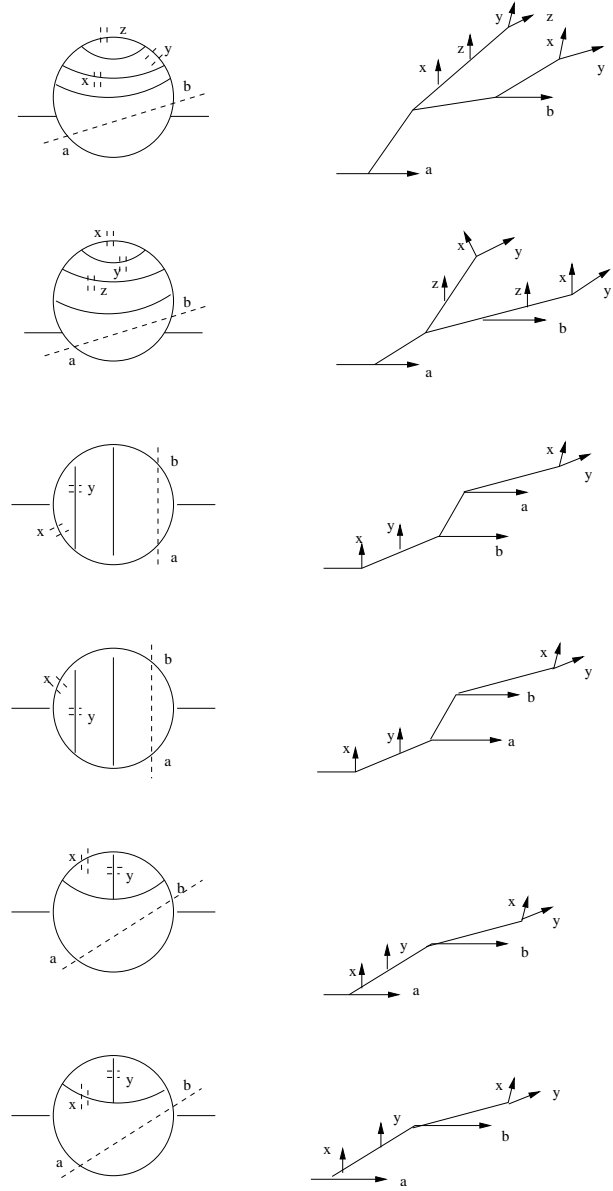}
\end{center}
\caption{Self energies with more than one type-B tic-ed propagator and the
corresponding scattering amplitudes 
}
\label{fgmn24}
\end{figure}

\subsubsection{Interpretation of the Rule}

The physical interpretation of these results is straightforward. Consider the
case of two tic-ed propagators. If both propagators were type-A, the thermal
factor would be determined by Eqn.(\ref{ticruleA}) which gives (up to numerical
factors),
\bea
N_1 N_2 = (e_1+a_1)(e_2+a_2)\nonumber
\eea
where $e_i=(1+n_i)$ is the statistical weight for the emission and $a_i=n_i$ is
the statistical weight for absorption.  This thermal factor tells us that
type-A fields are spectators whose emission and absorption is uncorrelated with
the emission and absorption of any of the other fields.

If both propagators were type-B, the thermal factor would be determined by
Eqn.(\ref{ticrule}) which gives (up to numerical factors and depending on
which way the momentum is routed),
\bea
e_1 e_2+a_1 a_2 \nonumber
\eea
This result tell us that the production of the fields associated with two
type-B tic-ed propagators is correlated, and that the two emitted-absorbed
pairs of fields interact with the system in a non trivial way. This result is
consistent with the information we obtain from the scattering amplitude: for
each of the two emitted-absorbed pairs, one of the partners appears on the
opposite side of the emitted/absorbed line produced by a cut propagator, and
thus they cannot be treated as spectator fields.
The plus sign between the two terms corresponds to the fact that emission and
absorption takes place on the same side of the cut line: we do not take the
difference of the weighting factors for a given process and its inverse, but
the sum of the two factors.  Note that the plus sign appears in exactly the
same way in Eqn.(\ref{ticruleA}): $N/2=1/2((1+n)+n)$.

In the case of one type-B propagator, and an arbitrary number of type-A
propagators, the lone type-B field does not have any other field with which to
correlate, and thus the factor given by Eqn.(\ref{ticruleA}) is the correct
one.

\subsection{Numerical Factors}

All diagrams carry an overall factor 
\bea
(2\pi)^{L+1} \frac{1}{(2\pi)^{4L}} \left(\int dp_i p_i^2 d\Omega_i\right)^L \label{OverAll} 
\eea
where $L$ is the number of loops. There is an additional factor, before thermal factors are considered, which is given by \cite{JohnS}
\bea
\left(\frac{1}{2}\right)^{p} 2^{v-1}i(i^p)(-i)^v \label{first}
\eea
where $p$ is the number of propagators and $v$ is the number of vertices.
The numerical factor contributed by the thermal factor is
\bea
2^{(C-1)} 2^{(CorrT-1)}2 (1/2)^{UnCorrT} \nonumber
\eea
where $C$ is the number of cut propagators, $CorrT$ is the number of correlated
tic-ed propagators or the total number of factors involved in the products over
$i$ and $j$ in Eqn.(\ref{ticrule}), and $UnCorrT$ is the number of uncorrelated
tic-ed propagators or the number of thermal factors of the form $N/2$ from
Eqn.(\ref{ticruleA}). The second two factors in this expression: $2
(1/2)^{UnCorrT}$ are explicitly included in the rules given by
Eqns.(\ref{ticruleA}), (\ref{ticrule}). The first two factors appear when one
rewrites the thermal factors that one obtains directly from the Keldysh RTF
Feynman rules in the physically motivated forms given by Eqns.(\ref{cutlines}),
(\ref{ticrule}). 
Combining these results and using 
\bea
&&C+CorrT + UnCorrT = L+1 \nonumber\\
&& L=p+1-v \nonumber 
\eea
we find that all factors of 2 cancel and, in addition to Eqn.(\ref{OverAll}), we are left with a numerical factor of

\bea
i(i^p)(-i)^v \nonumber
\eea

\section{Examples}

In this section we give a list of the diagrams we calculated in order to deduce
the rules described above. The calculations were performed using the {\it
Mathematica} program in \cite{JohnS}.

\subsection {Two Loop Self Energy Formed by a Propagator Correction}

The two loop self energy that is formed from the one loop self energy by adding
a correction to one of the propagators is shown in Fig.[3]. There are five
cuts, as shown in Fig.[4a-e]. The cut in Fig.[4e] produces three cut
propagators and carries a thermal factor $(1+n_{p+q+k})n_k n_p -
n_{p+q+k}(1+n_p)(1+n_k)$ according to rule in Eqn.(\ref{cutlines}). The graphs
in Figs.[4a-d] contain two cut propagators and one tic-ed propagator. Following
the rules given in Eqns. (\ref{cutlines}), (\ref{ticruleA}) the thermal factor
for Figs.[4a,c] is $N_k((1+n_{p+q})n_p-n_{p+q}(1+n_p))$. In the same way, the
thermal factor for Figs.[4b,d] is $N_{k+q+p}((1+n_{p+q})n_p-n_{p+q}(1+n_p))$.
For a scalar theory, the diagrams in Figs.[4a,d] can be written in the same
form by shifting momentum integrals, and/or taking the complex conjugate. For any theory with a cubic interaction,
these diagrams have the same structure. In the future, in order to compactify
the notation,  we will represent a set of diagrams of this form by drawing one diagram and writing a factor 
$\otimes N$ next to it, to indicate that there are $N$ permutations of the
diagram.
\begin{figure}[H]
\begin{center}
\includegraphics[width=8cm]{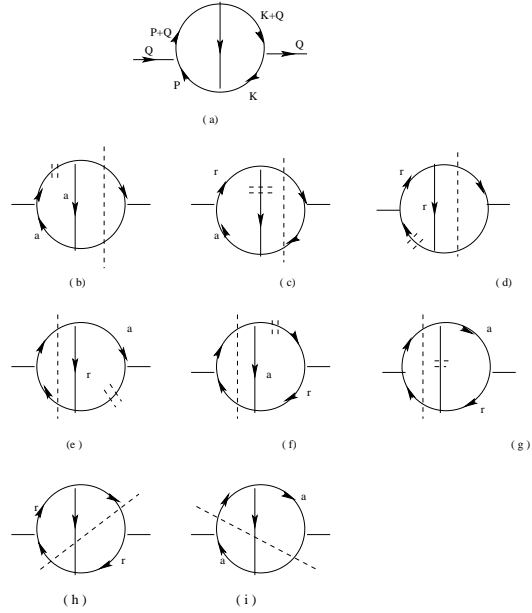}
\end{center}
\caption{A two loop self energy and its cuts
}
\label{fgm15}
\end{figure}


\subsection {Two Loop Self Energy with a Vertex Correction}

The two loop self energy that is formed from the one loop self energy by adding
a correction to one of the vertices is shown in Fig.[26a]. There are eight
cuts, as shown in Fig.[26b-i].  The thermal factors are obtained from the rules
given in Eqns.(\ref{cutlines}), (\ref{ticruleA}). Figs.[26h,i] have three cut
propagators and carry the thermal factors $(1+n_{q+p})n_{p-k}n_k -
n_{p+q}(1+n_{p-k})(1+n_k)$ and $(1+n_{q+k})(1+n_{p-k})n_p -
n_{k+q}n_{p-k}(1+n_p)$, respectively. Figs.[26b-g] have two cut propagators and
one tic-ed propagator and carry the factors:
\bea
&&[(1+n_{k+q})n_k-n_{k+q}(1+n_k)]N_{p+q}/2\,;\nonumber \\
&&[(1+n_{k+q})n_k-n_{k+q}(1+n_k)]N_{p-k}/2\,;\nonumber \\
&& [(1+n_{k+q})n_k-n_{k+q}(1+n_k)]N_p/2\,;\nonumber \\
&&[(1+n_{p+q})n_p-n_{p+q}(1+n_p)]N_k/2\,;\nonumber \\
 &&[(1+n_{p+q})n_p-n_{p+q}(1+n_p)]N_{k+q}/2\,;\nonumber \\
&&[(1+n_{p+q})n_p-n_{p+q}(1+n_p)]N_{p-k}/2\,.\nonumber
\eea
 In the future, we will compactify the notation by noting that
Figs.[26b,d,e,f], and  Figs.[26c,g] are permutations of each other. We will
draw Fig.[26b], with a factor $\otimes 4$ and Fig.[26c], with a factor $\otimes
2$.

\subsection{Further Examples}

The thermal factors for the diagrams listed below are determined from the rules
given by Eqns.(\ref{cutlines}), (\ref{ticruleA}) and (\ref{ticrule}). To
simplify notation, we define the symbol
\bea
{\cal N}(x_1,\cdots x_n;y_1,\cdots y_m) = [(1+n(x_1)\cdots(1+x(x_n)n(y_1)\cdots
y(x_n)-n(x_1)\cdots n(x_n)(1+n(y_1))\cdots(1+n(y_m))]\nonumber
\eea

\par\begin{figure}[H]
\begin{center}
\includegraphics[width=8cm]{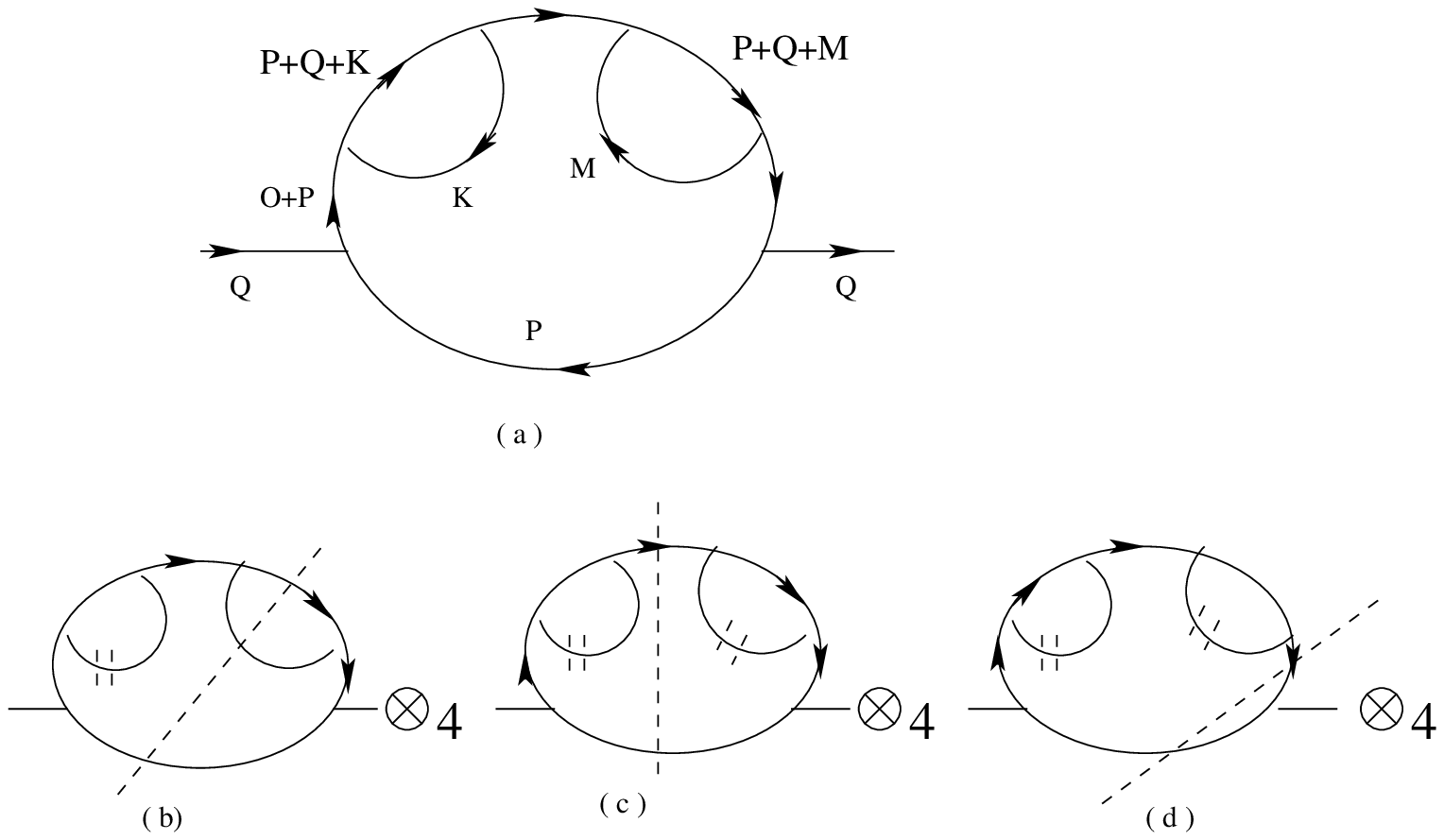}
\end{center}
\caption{A three loop self energy and its cuts.
}
\label{3l2hn}
\end{figure}
The thermal factors for Figs.[27b-d] are respectively,
\bea
&&\frac{1}{2}N(k){\cal N}(p+q+m;m,p) \nonumber \\
&&\frac{1}{4}N(k)N(m){\cal N}(p+q;p)\nonumber \\
&&\frac{1}{4}N(m)N(k){\cal N}(p+q;p)\nonumber 
\eea
\par\begin{figure}[H]
\begin{center}
\includegraphics[width=8cm]{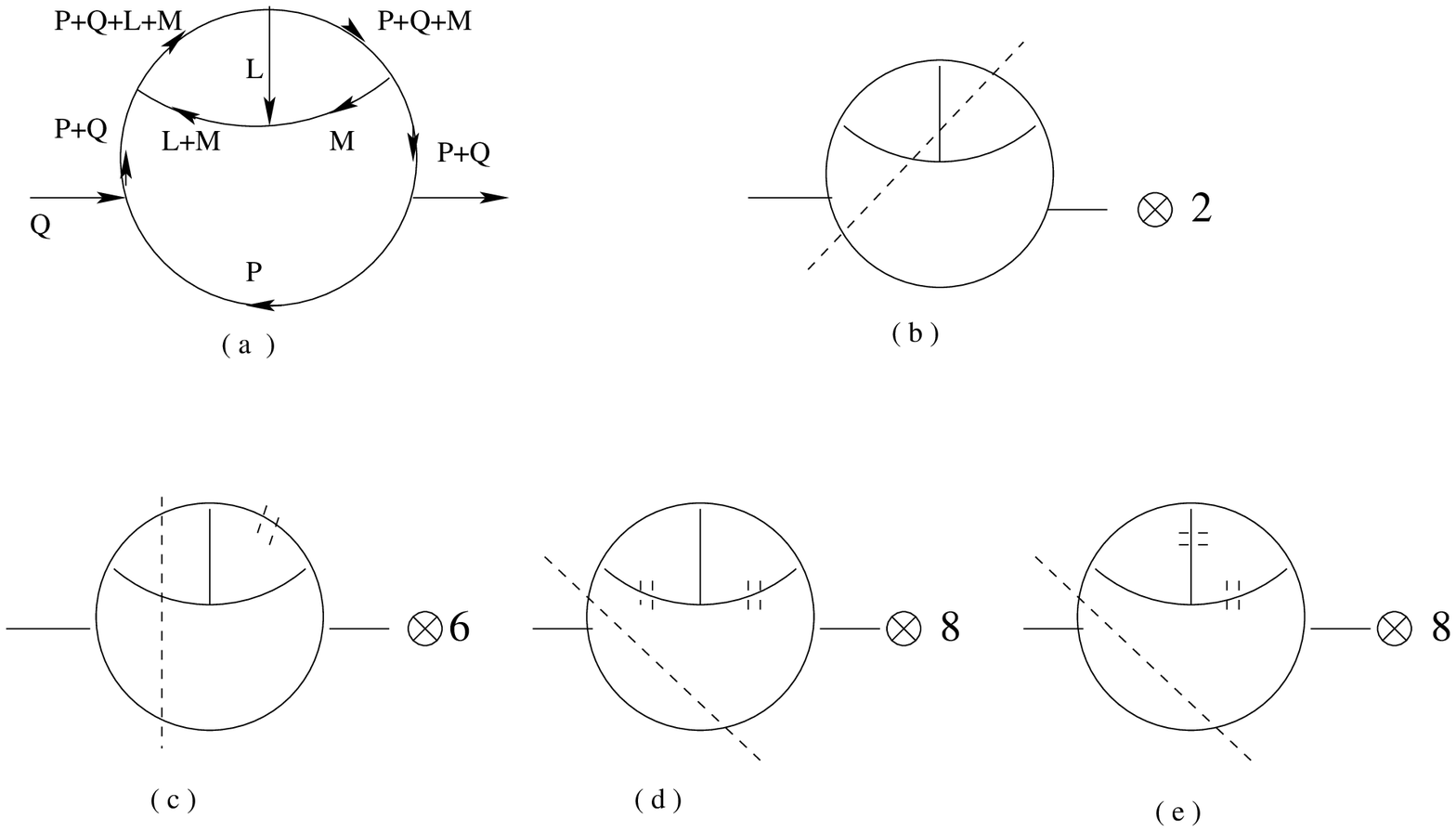}
\end{center}
\caption{A three loop self energy and its cuts.
}
\label{3leye}
\end{figure}
The thermal factors for Figs.[28b-e] are respectively,
\bea
&&{\cal N}(p+q+m,l;p,l+m)\nonumber \\
&&\frac{1}{2}N(p+q+m){\cal N}(p+q+l+m;p,l+m)\nonumber \\
&&\frac{1}{4} N(m) N(l+m){\cal N}(p+q;p)\nonumber \\
&& \frac{1}{4}N(l) N(m){\cal N}(p+q;p) \nonumber
\eea
\par\begin{figure}[H]
\begin{center}
\includegraphics[width=8cm]{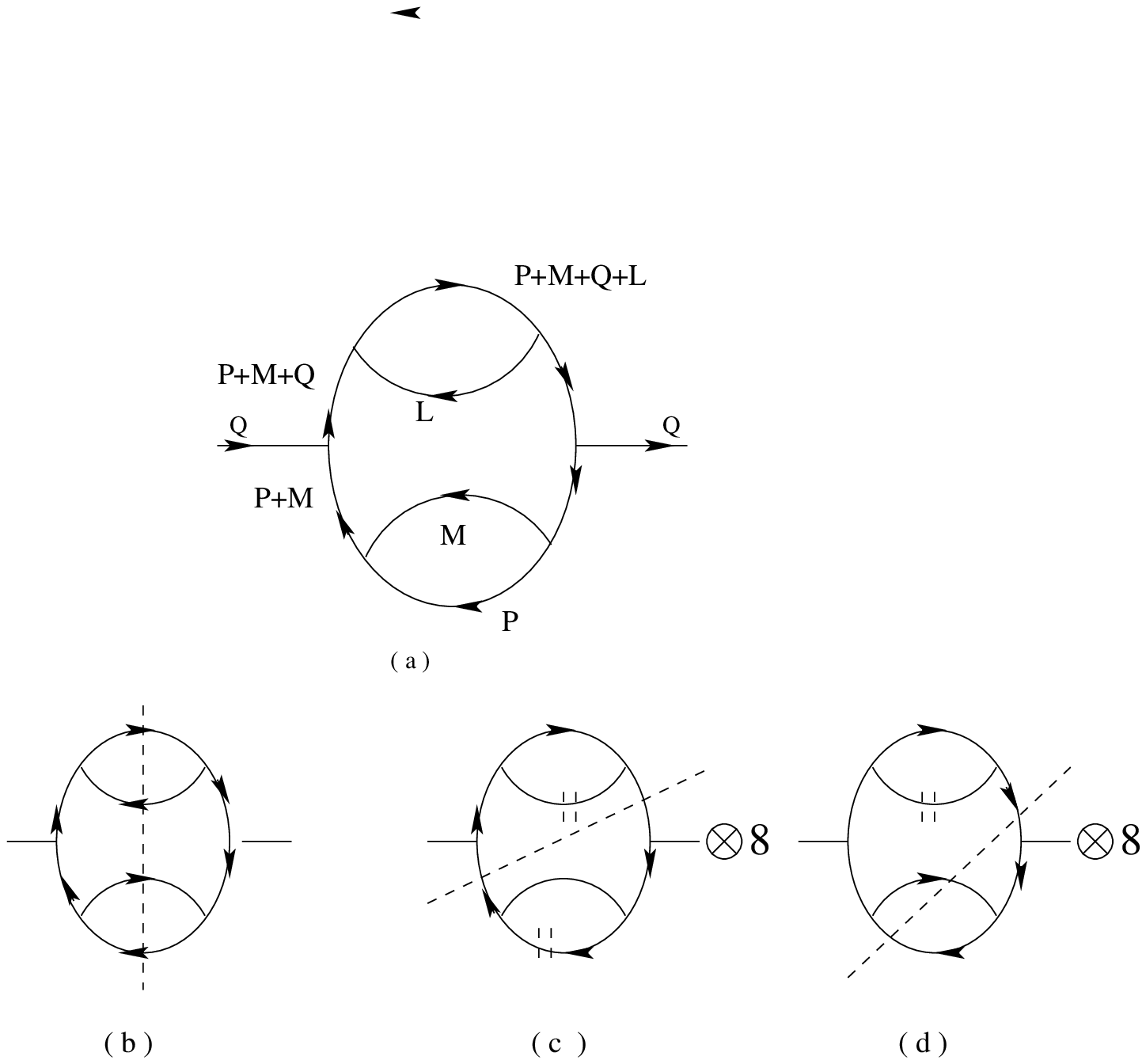}
\end{center}
\caption{A three loop self energy and its cuts.
}
\label{3lh2n}
\end{figure}
The thermal factors for Figs.[29b-d] are respectively,
\bea
&&{\cal N}(p+q+m+l;p,m,l)\nonumber \\
&&\frac{1}{4}N(l)N(p){\cal N}(p+q+m;p+m) \nonumber \\
&&\frac{1}{2}N(l){\cal N}(p+q+m;p,m) \nonumber
\eea
\par\begin{figure}[H]
\begin{center}
\includegraphics[width=8cm]{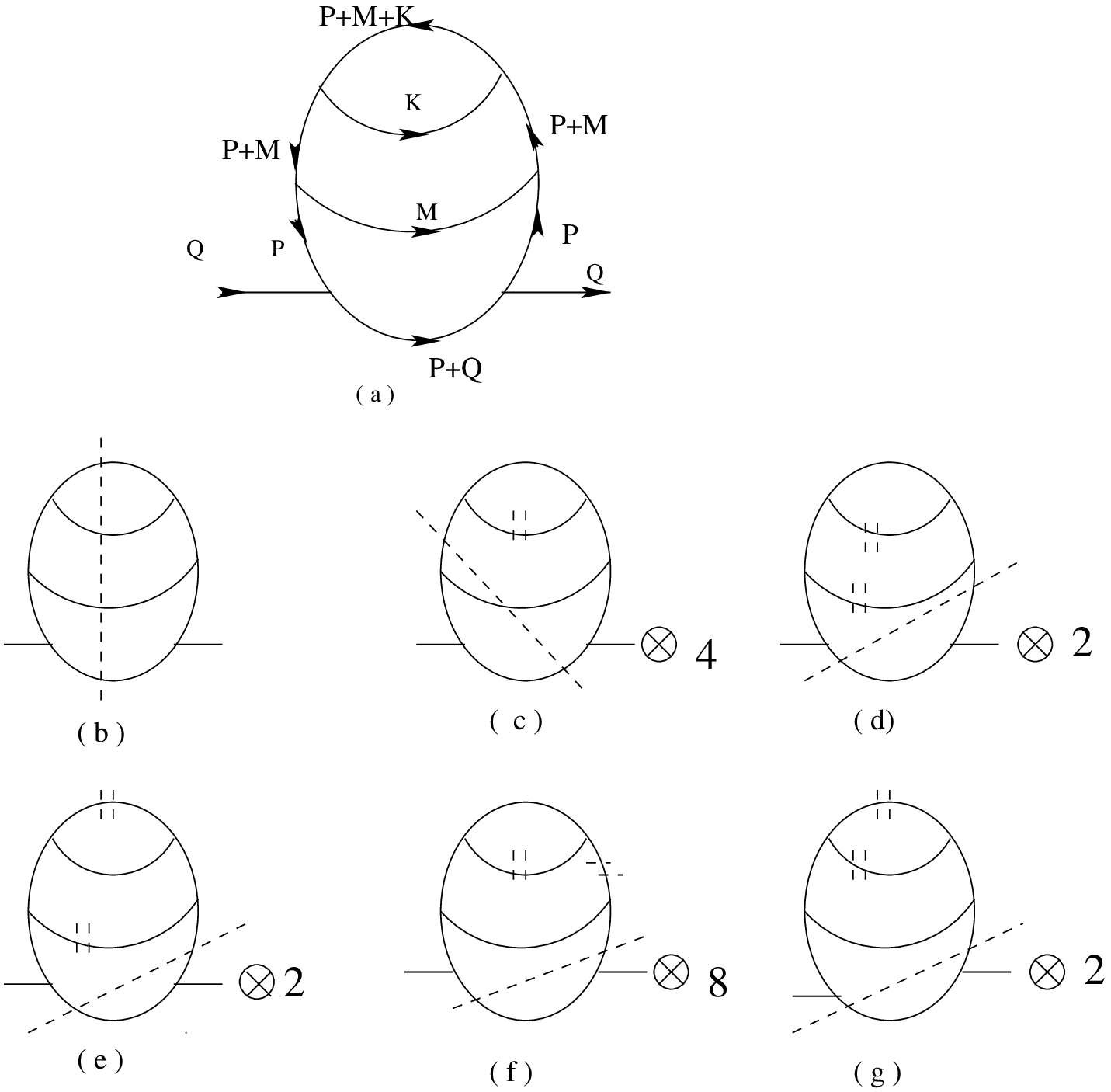}
\end{center}
\caption{A three loop self energy and its cuts.
}
\label{3l2sn}
\end{figure}
The thermal factors for Figs.[30b-g] are respectively,
\bea
&&{\cal N}(p+q,m,k;p+m+k)\nonumber \\
&&\frac{1}{2}N(k){\cal N}(p+q,m;p+m) \nonumber \\
&&\frac{1}{4}N(k)N(m){\cal N}(p+q;p)\nonumber \\
&&\frac{1}{4}N(m)N(p+m+k){\cal N}(p+q;p)\nonumber \\
&&\frac{1}{4}N(k)N(p+m){\cal N}(p+q;p)\nonumber \\
&&2[(1+n_k)n_{p+m+k}+n_k(1+n_{p+m+k})]{\cal N}(p+q;p)\nonumber 
\eea
\par\begin{figure}[H]
\begin{center}
\includegraphics[width=8cm]{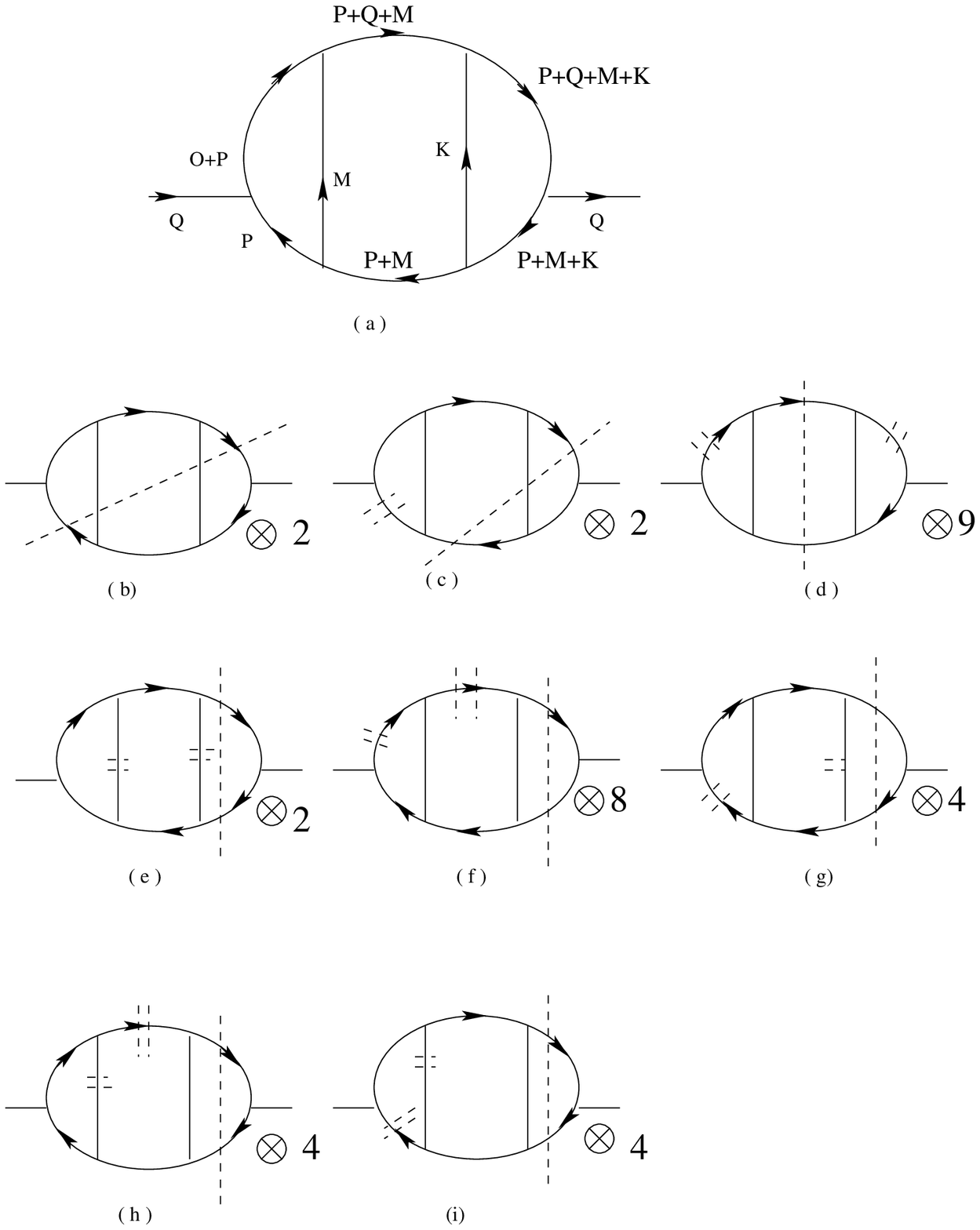}
\end{center}
\caption{A three loop self energy and its cuts.
}
\label{3lvnn}
\end{figure}
The thermal factors for Figs.[31b-i] are respectively,
\bea
&&{\cal N}(p+q+m+k;k,m,p)\nonumber \\
&&\frac{1}{2}N(p){\cal N}(p+q+m+k;k,p+m)\nonumber \\
&& \frac{1}{4}N(q+p)N(q+p+m+k){\cal N}(p+q+m;p+m)\nonumber \\
&&\frac{1}{4}N(m)N(k){\cal N}(p+q+m+k;p+m+k)\nonumber\\
&&\frac{1}{4}N(p+q)N(p+q+m){\cal N}(p+q+m+k;p+m+k)\nonumber\\
&&\frac{1}{4}N(p)N(k){\cal N}(p+q+m+k;p+m+k)\nonumber\\
&&\frac{1}{4}N(m)N(p+q+m){\cal N}(p+q+m+k;p+m+k)\nonumber\\
&&2[(1+n_m)(1+n_p)+n_m n_p]{\cal N}(p+q+m+k;p+m+k)\nonumber
\eea

\par\begin{figure}[H]
\begin{center}
\includegraphics[width=10cm]{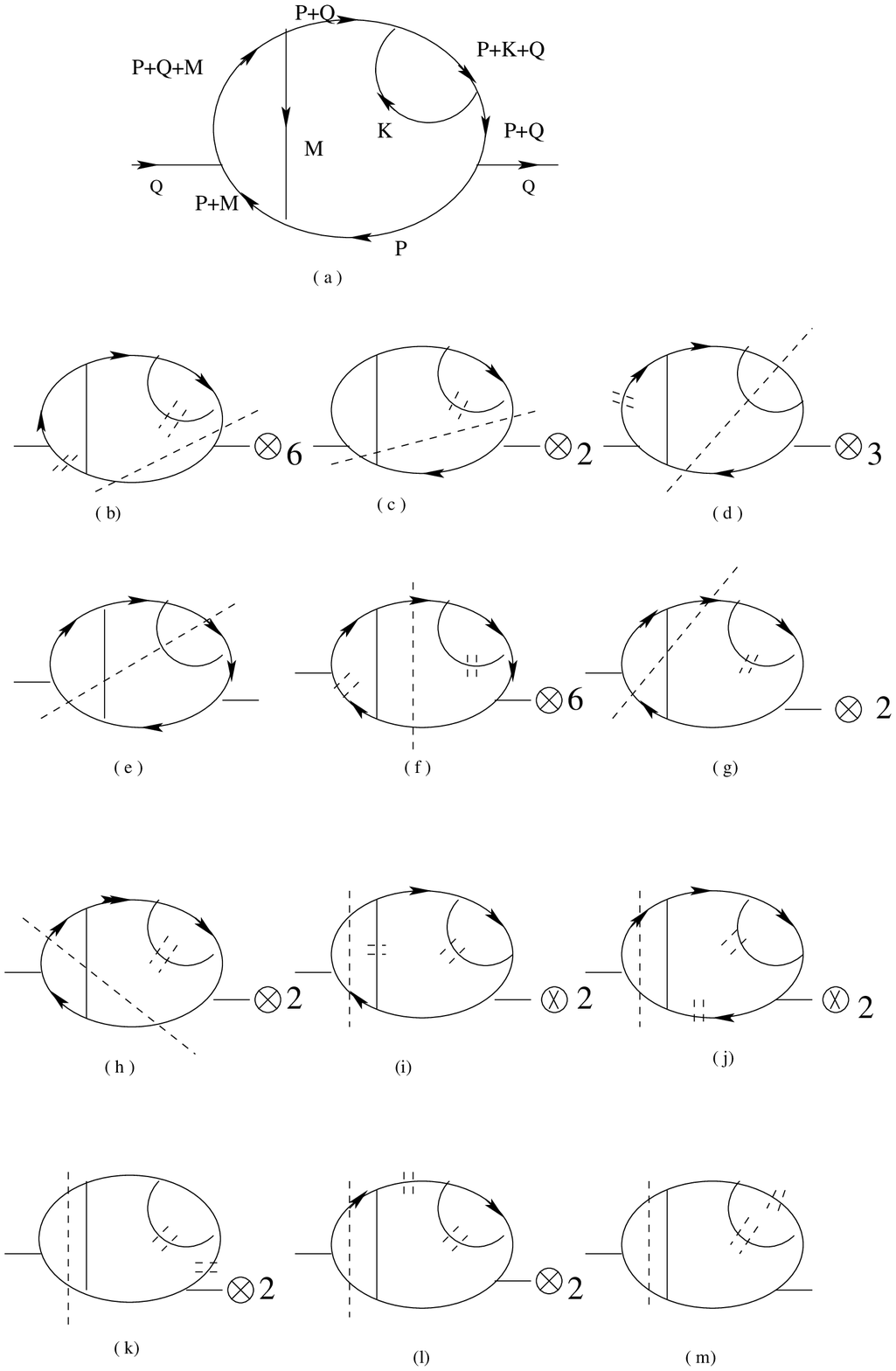}
\end{center}
\caption{A four loop self energy and its cuts.
}
\label{3lvhnn}
\end{figure}


The thermal factors for Figs.[32b-m] are respectively,
\bea
&&\frac{1}{4}N(k)N(p+m){\cal N}(p+q;p)\nonumber \\
&&\frac{1}{2}N(k){\cal N}(p+q,m;p+m)\nonumber \\
&&\frac{1}{2}N(p+m+q){\cal N}(p+q+k;p,k)\nonumber \\
&&{\cal N}(p+q+k,m;p+m,k)\nonumber \\
&&\frac{1}{4}N(k)N(m){\cal N}(p+q;p)\nonumber\\
&&\frac{1}{2}N(k){\cal N}(p+q,m;p+m)\nonumber \\
&&\frac{1}{2}N(k){\cal N}(p+q+m;p,m)\nonumber \\
&&\frac{1}{4}N(k)N(m){\cal N}(p+q+m;p+m)\nonumber \\
&&\frac{1}{4}N(k)N(p){\cal N}(p+q+m;p+m)\nonumber \\
&&\frac{1}{4}N(k)N(p+q){\cal N}(p+q+m;p+m)\nonumber \\
&&\frac{1}{4}N(k)N(p+q){\cal N}(p+q+m;p+m)\nonumber \\
&&2[(1+n_k)n_{q+p+k}+n_k(1+n_{q+p+k})]{\cal N}(p+q+m;p+m)\nonumber 
\eea
\par\begin{figure}[H]
\begin{center}
\includegraphics[width=15cm]{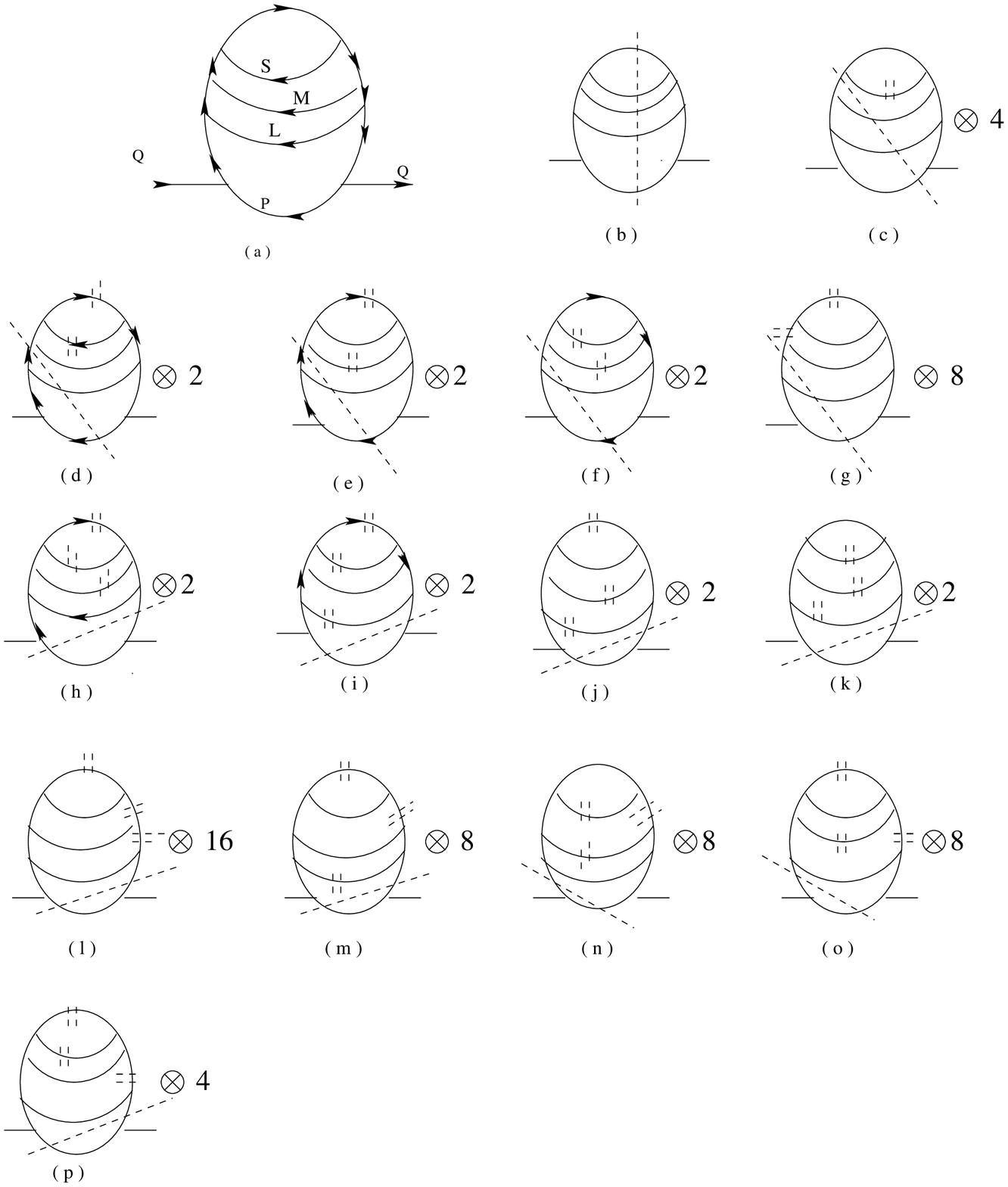}
\end{center}
\caption{A four loop self energy and its cuts.
}
\label{4l3sn}
\end{figure}
The thermal factors for Figs.[33b-p] are respectively,
\bea
&&{\cal N}(p+q+m+l+s;s,m,l,p)\nonumber \\
&&\frac{1}{2}N(s){\cal N}(p+q+m+l;m,l,p)\nonumber \\
&&2[(1+n_s)n_{p+q+m+l+s}+n_s(1+n_{p+q+m+l+s})]{\cal N}(p+q+l;l,p)\nonumber \\
&&\frac{1}{4}N(p+q+m+l+s)N(m){\cal N}(p+q+l;l,p)\nonumber \\
&&\frac{1}{4}N(s)N(m){\cal N}(p+q+l;l,p)\nonumber \\
&&\frac{1}{4}N(p+q+m+l+s)N(p+q+l+m){\cal N}(p+q+l;l,p)\nonumber \\
&&2[(1+n_{p+q+m+l+s}n_s n_m + n_{p+q+m+l+s}(1+n_s)(1+n_m)]{\cal
N}(p+q;p)\nonumber \\
&& N(l)[(1+n_{p+q+m+l+s})n_s+n_{p+q+m+l+s}(1+n_s)]{\cal N}(p+q;p)\nonumber \\
&&\frac{1}{8}N(p+q+m+l+s)N(m)N(l){\cal N}(p+q;p)\nonumber \\
&&\frac{1}{8}N(s)N(m)N(l){\cal N}(p+q;p)\nonumber \\
&&\frac{1}{8}N(p+q+m+l+s)N(p+q+m+l)N(p+q+l){\cal N}(p+q;p)\nonumber \\
&&\frac{1}{8}N(p+q+m+l+s)N(p+q+m+l)N(l){\cal N}(p+q;p)\nonumber \\
&&N(s)[(1+n_{p+q+m+l}n_m+n_{p+q+m+l}(1+n_m)]{\cal N}(p+q;p)\nonumber \\
&&\frac{1}{8}N(m)N(p+q+m+l+s)N(p+q+l){\cal N}(p+q;p)\nonumber \\
&&N(p+q+l)[(1+n_{p+q+m+l+s})n_s+n_{p+q+m+l+s}(1+n_s)]{\cal N}(p+q;p)\nonumber 
\eea

\vspace*{2cm}

\section{Conclusions}

In this paper we have discussed a set of rules for calculating the imaginary
parts of self energy diagrams as a
series of tree amplitudes which represent physical scattering processes. The
thermal factors associated with each scattering diagram have a physical
interpretation in terms of the statistical weighting factors associated with
the emission and absorption of thermal fields. Work on a general
derivation of these rules from first principles is in progress.

\end{document}